%% file: main.tex
\documentclass{jpp}
\usepackage{graphicx, subcaption}
\usepackage{epstopdf}

\usepackage[utf8]{inputenc}
\usepackage[T1]{fontenc}\usepackage{amsmath,amsfonts,amssymb}
\usepackage{cancel,braket}
\usepackage{caption}
\usepackage{float}
\usepackage{slashed}
\usepackage{bbold}
\usepackage{hyperref}
\usepackage{tcolorbox} 
\usepackage{enumitem} 
\usepackage{schemata}
\usepackage{enumerate}
\usepackage{multirow}
\usepackage{pdfpages}
\usepackage{longtable}
\usepackage{parcolumns}
\usepackage{physics}
\usepackage{booktabs}
\usepackage{xcolor}
\usepackage{amsmath}
\usepackage{listings}

\title{
{Electrostatic gyrokinetic simulations in Wendelstein 7-X geometry: benchmark  between the codes \texttt{stella} and \texttt{GENE}}
}

\author{A. González-Jerez\aff{1}
  \corresp{\email{AntonioG@ciemat.es}},
  P. Xanthopoulos\aff{2}, J. M. García-Regaña\aff{1}, \\I. Calvo\aff{1},
  J. Alcusón\aff{2}, A. Bañón-Navarro\aff{3},  M. Barnes\aff{4},\\
  F. I. Parra\aff{4} \and J. Geiger\aff{2}}

\shorttitle{Benchmark of {stella} against {GENE} in W7-X geometry}
\shortauthor{A. González-Jerez et al.}

\affiliation{\aff{1}Laboratorio Nacional de Fusión, CIEMAT, 28040, Madrid, Spain
\aff{2}Max-Planck Institut für Plasmaphysik, 17491 Greifswald, Germany 
\aff{3}Max-Planck Institut für Plasmaphysik, 85748 Garching, Germany  
\aff{4}Rudolf Peierls Centre for Theoretical Physics, University of Oxford, Oxford OX1 3PU, UK}

\shorttitle{Benchmark of stella against GENE in W7-X geometry}
\shortauthor{A. González-Jerez et al.}

\begin{document}

\maketitle

\input{files/abstract}

\input{files/Introduction}

\input{files/equations}

\input{files/parameters}

\input{files/Linear}

\input{files/Non_linear}

\input{files/Summary}

\input{files/Acknow}

\appendix
\input{files/AP_1}

\bibliographystyle{jpp}
\bibliography{Bench_paper.bib}

\end{document}

%% file: files/abstract.tex
\begin{abstract}


The first experimental campaigns have proven that, due to the optimization of the magnetic configuration with respect to neoclassical transport, the contribution of turbulence is essential to understand and predict the total particle and energy transport in Wendelstein 7-X (W7-X). This has spurred much work on gyrokinetic modelling for the interpretation of the available experimental results and for the preparation of the next campaigns. At the same time, new stellarator gyrokinetic codes have just been or are being developed. It is therefore desirable to have a sufficiently complete, documented and verified set of {gyrokinetic simulations in W7-X geometry against} which new codes or upgrades of existing codes can be tested and benchmarked. This paper {attemps to provide such a set of simulations in the form of a comprehensive benchmark between the recently developed code \texttt{stella} and the well-established code \texttt{GENE}. The benchmark consists of electrostatic gyrokinetic simulations in W7-X magnetic geometry and includes different flux tubes, linear {ion-temperature-gradient (ITG) and trapped-electron-mode (TEM)} stability analyses, computation of linear zonal flow responses and calculation of ITG-driven heat fluxes.}

\end{abstract}

%% file: files/Introduction.tex
\section{Introduction}

One of the main issues in magnetic confinement fusion plasmas is the theoretical understanding of turbulence and turbulent transport, attributable to the action of instabilities driven by density and temperature gradients. Since decades, turbulence is known to be the main source of transport in low collisionality tokamak plasmas. In stellarators, neoclassical transport at low collisionality is usually large, at least in the plasma core, and turbulence has {sometimes} been assumed to play a less important role. The first experimental campaigns of Wendelstein 7-X (W7-X) (\citet{wolf2017}; \citet{Klinger2019}) have made it evident that, in general, the total particle and energy transport is higher than predicted by neoclassical theory \citep{Bozhenkov2020}, turning turbulent transport into an essential mechanism to understand and predict these results. This disagreement between the experimental measurements and the predictions of neoclassical theory has motivated much work on gyrokinetic modelling, including upgrades of existing codes and development of new ones with the aim of understanding the first available experimental results of W7-X and preparing the next experimental {campaigns}.

{
While existing tokamak gyrokinetic codes (\citet{Parker1993}, \citet{Kotschenreuther1995}, \citet{Lin1998}, {\citet{Dorland2000}, \citet{Jenko2000}}, \citet{Candy2003}, \citet{Jolliet2007}, \citet{Peeters2009}) have been extensively exploited and tested, less work has been carried out in the validation and verification of stellarator gyrokinetic codes (\citet{Kornilov2004}, \citet{Watanabe2005}, \citet{Xanthopoulos2007_linear},  \citet{Baumgaertel2011}, \citet{Cole2019}, \citet{Barnes2019}, \citet{Maurer2020}, \citet{Wang2020}).
} In this context, it is worthwhile to have a sufficiently complete, documented and well verified set of linear and nonlinear gyrokinetic simulations in W7-X {geometry against which present and future stellarator gyrokinetic codes can be tested and benchmarked (following the example of the tokamak community with the Cyclone Base Case \citep{Dimits2000}).} This paper is an attempt to provide {such a set of simulations in the form of a comprehensive benchmark between the recently developed flux-tube gyrokinetic code \texttt{stella} and the well-established code \texttt{GENE} in W7-X geometry}. The most important difference between these two codes is the treatment of the parallel streaming and acceleration terms of the gyrokinetic Vlasov equation. The mixed implicit-explicit numerical scheme used by \texttt{stella} makes it possible to handle these terms implicitly, allowing a larger time step size in simulations with kinetic electrons. This turns \texttt{stella} into an efficient code for multispecies turbulence simulations in stellarators.

 The rest of the paper is organized as follows. In section \ref{sec.codes}, the flux tube equations solved by \texttt{stella} and \texttt{GENE} are presented, as well as some relevant differences between their implementation in these two codes. In section \ref{sec.parameters}, the W7-X magnetic configuration selected for our study is described together with the two flux tubes in which the simulations will be performed, the so-called \textit{bean} and \textit{triangular} flux tubes. The simulations of this paper are divided into five \textit{tests} and the parameters used to carry out each one of them are also collected in this section. In section \ref{sec.lineal}, the linear part of the study is performed, encompassing the first four tests. In tests 1 to 3, the values of growth rate and {real} frequency computed with \texttt{GENE} and \texttt{stella} are compared. Tests 1 and 2 assume adiabatic electrons, studying ITGs in the bean and triangular flux tubes, and test 3 includes kinetic electrons, studying {density-gradient-driven} TEMs in the bean flux tube. In these tests, the structure of the electrostatic potential is also {given}, discussing the features of each instability. The remarkable difference between the time step size required by \texttt{stella} and \texttt{GENE} in linear simulations with kinetic electrons is also emphasized. In test 4, the linear zonal flow response is computed with both codes, comparing four different time traces of the electrostatic potential relaxation. In section \ref{sec.5}, nonlinear simulations results are given. They include the study of the ITG-driven heat flux and the contribution of each mode to this quantity, computed with \texttt{stella} in the bean and triangular flux tubes. The results obtained in the bean flux tube are compared with \texttt{GENE} calculations in test 5. {Finally, section 6 {contains the summary and the conclusions.}}

%% file: files/equations.tex
\section{Equations solved by \texttt{stella} and \texttt{GENE}}\label{sec.codes}

$\texttt{GENE}$ and $\texttt{stella}$ are based on the $\delta f$-gyrokinetic theory, first proposed in \citet{Catto1978}. They solve the gyrokinetic Vlasov and quasineutrality equations for an arbitrary number of species. In this work, both codes solve the flux tube version of these equations, which we explicitly write below. The spatial coordinates are denoted by $\{x,y,z\}$, where $x \in [0,a]$ is a radial coordinate that labels magnetic surfaces with $a$ the minor radius of the device, $y \in [0,2\pi)$ is an angular coordinate labeling field lines on each magnetic surface and $z\in [z_{\mathrm{min}},z_{\mathrm{max}}]$ is a coordinate along magnetic field lines. Using this notation and assuming nested flux surfaces, the magnetic field can be expressed as $\mathbf{B}=\psi'_t\nabla x\times \nabla y$, where a prime $(')$ means differentiation with respect to the coordinate $x$ and $2\pi\psi_t$ is the toroidal flux. The velocity coordinates employed throughout this paper are $\{v_{||},\mu\}$, where $v_{||}$ is the component of the velocity parallel to the magnetic field line and $\mu=m_jv_{\perp}^2/2B$ is the magnetic moment, with  $v_{\perp}$ the component of the velocity perpendicular to the magnetic field, $B$ the magnetic field strength and $m_j$ the mass, where the subscript $j$ indicates the species. In this work $j$ can take the values `$i$', if it refers to ions, and `$e$', if it refers to electrons. Using these definitions, the flux tube gyrokinetic Vlasov equation for the mode $\mathbf{k}$ of the fluctuating distribution function, $\hat{g}_{\mathbf{k},j}$, reads
\begin{align}\label{norm_gyro_eq}
     \partial_t\hat{g}_{\mathbf{k},j}
     &
     +v_{\|}\mathbf{\hat{b}}\cdot\nabla z\left(\partial_z\hat{g}_{\mathbf{k},j}+ \frac{Z_je}{T_j}\partial_{z}\left[\hat{\varphi_{\mathbf{k}}}J_0(k_{\perp}\rho_j)\right]F_{0,j}\right) 
     -\frac{\mu}{m_j}\mathbf{\hat{b}}\cdot\nabla B\partial_{v_{\|}}\hat{g}_{\mathbf{k},j}
     \nonumber
     \\
     & 
    -\mathrm{i} \frac{k_{y}}{\psi'_t}\left[\frac{n'_j}{n_j}+\frac{T'_j}{T_j}\left(\frac{m_j(v_{\|}^2/2+B\mu/m_j)}{T_j}-\frac{3}{2}\right)\right]\hat{\varphi}_{\mathbf{k}}J_0(k_{\perp}\rho_j)F_{0,j}
     \\
     & 
    +\frac{\mathrm{i}}{\Omega_{j}}\left(v_{\|}^2\mathbf{\hat{b}}\times\boldsymbol{\kappa}+\frac{\mu}{m_j}\mathbf{\hat{b}}\times\nabla B\right)\cdot\mathbf{k_{\perp}}\left(\hat{g}_{\mathbf{k},j}+\frac{Z_je}{T_j}\hat{\varphi}_{\mathbf{k}}J_0(k_{\perp}\rho_j)F_{0,j}\right) - N_{\mathbf{k},j} =0,
    \nonumber
\end{align}
 where $t$ is the time, $n_j(x)$ and $T_j(x)$ are the density and temperature, $\rho_j=v_{\mathrm{th},j}/\Omega_j$ is the gyroradius, $v_{\mathrm{th},j}=\sqrt{2T_j/m_j}$ is the thermal speed, $\Omega_j=Z_jeB/m_j$ is the gyrofrequency, $Z_j$ is the charge number, $e$ is the proton charge, $\mathbf{k_{\perp}}=k_{x}\nabla x + k_{y}\nabla y$ is the perpendicular wavevector,
 \begin{equation*}
     F_{0,j}=n_j(m_j/2\pi T_j)^{3/2}\exp[-m_j(v_{||}^2/2 + \mu B/m_j)/T_j]
 \end{equation*} 
 is a Maxwellian and $J_0$ is the zeroth order Bessel function of the first kind. Finally, $\hat{\varphi}_{\mathbf{k}}$ is the mode $\mathbf{k}$ of the fluctuating electrostatic potential, $\boldsymbol{\kappa}=\mathbf{\hat{b}}\cdot\nabla\mathbf{\hat{b}}$ is the curvature vector, with $\mathbf{\hat{b}}=B^{-1}\mathbf{B}$ and $N_{\mathbf{k},j}$ is the mode $\mathbf{k}$ of the nonlinear term, which can be written as
\begin{equation*}
    N_{{\mathbf{k}},j} =\frac{1}{B} \sum_{\underset{k_{x_2}, k_{y_2}}{k_{x_1}, k_{y_1}}}\mathbf{\hat{b}}\cdot\left(\mathbf{k}_{\perp_1}\times \mathbf{k}_{\perp_2}\right)J_0\left(k_{\perp_1}\rho_j\right)\hat{\varphi}_{\mathbf{k}_1}\hat{g}_{\mathbf{k}_2,j},
\end{equation*}
   where $\mathbf{k}_{\perp_1}$ and $\mathbf{k}_{\perp_2}$ are such that $\mathbf{k}_{\perp_1} + \mathbf{k}_{\perp_2} =\mathbf{k}_{\perp}$.
   
   The quasineutrality equation reads
\begin{align}\label{quasi}
    \sum_j\frac{Z_jB}{m_j}\int_{-\infty}^{\infty}\dd v_{\|}\int_0^{\infty}\dd\mu
    J_0(k_{\perp}\rho_j)\hat{g}_{\mathbf{k},j}
    +\sum_j \frac{Z_j^2 e n_j}{2\pi T_j}\left(\Gamma_0\left(k_{\perp}^2\rho_j^2/2\right)-1\right)\hat{\varphi}_{\mathbf{k}} = 0,
\end{align}
where $\Gamma_0(b)=\exp(-b)I_0(b)$ and $I_0$ is the zeroth order modified Bessel function of the first kind.

Despite solving the same equations (\ref{norm_gyro_eq}) and (\ref{quasi}), there are some differences between \texttt{stella} and \texttt{GENE} when implementing them. In this paper only the most relevant ones are mentioned, see \citet{Barnes2019} and \citet{Merz_thesis} for further details. Both codes read the geometric quantities required to solve the Vlasov gyrokinetic and quasineutrality equations from a VMEC \citep{Hirshman1983} output, but they use different spatial coordinates. While the equations (\ref{norm_gyro_eq}) and (\ref{quasi}) are expressed in general flux coordinates $\{x,y,z\}$, \texttt{stella} and \texttt{GENE} use the  sets $\{a\sqrt{s}, a\alpha\sqrt{s_0},\zeta\}$ and $\{a\sqrt{s}, a\alpha\sqrt{s_0},\theta\}$, respectively, where $s$ is the toroidal magnetic flux normalized to its value on the last closed flux surface, $\alpha$ is the Clebsch angle and $\theta$ and $\zeta$ are, respectively, the poloidal and toroidal PEST flux coordinates \citep{Grimm1983}, related by $\alpha=\theta-\iota\zeta$, where $\iota$ is the rotational transform. Nevertheless, the main difference between \texttt{stella} and \texttt{GENE} is the treatment of the parallel streaming and the acceleration terms in equation (\ref{norm_gyro_eq}). For electrons, these terms of the gyrokinetic Vlasov equation scale up to a factor $\sqrt{m_i/m_e}$ with respect to the other terms. This imposes severe restrictions on the time step size in explicit methods when performing simulations with kinetic electrons. \texttt{GENE} treats these terms explicitly, while \texttt{stella} computes them implicitly, allowing to handle kinetic electrons using a time step size only slightly smaller than the one employed in simulations with adiabatic electrons, greatly reducing the computational cost. This will be clearly shown in section \ref{sec.lineal}, where simulations with kinetic electrons are presented. Finally, {different} boundary conditions have been used by each code in linear simulations. {Whereas} \texttt{stella} {employs a zero incoming boundary condition on the fluctuating distribution {function, $\hat{g}_{\mathbf{k},j}(z_{\mathrm{min}}, v_{\|}>0) = \hat{g}_{\mathbf{k},j}(z_{\mathrm{max}}, v_{\|}<0) = 0$}, \texttt{GENE} uses} \textit{twist and shift} boundary conditions \citep{Beer1995}.

%% file: files/parameters.tex
\section{Configuration and parameters}\label{sec.parameters}

The flexible set of planar and non-planar coils of W7-X allows a large variety of magnetic configurations. Depending on the values that some quantities take (by appropriately choosing the currents in the coil system), different names are used for each configuration. For instance, the value of $\iota$ allows to distinguish between \textit{high-} and \textit{low-iota} configurations, in which $\iota$ at the boundary is 5/4 and 5/6, respectively, and the magnetic configuration naturally develops the corresponding island structures. Similarly, depending on the toroidal mirror ratio\footnote{The toroidal mirror ratio is defined as $b_{01}\equiv B_{01}/B_{00}$ in a Fourier representation $B_{mn}$ of the magnetic field strength, where $m$ and $n$ are, respectively, poloidal and toroidal mode numbers. In different configurations of W7-X this ratio can range, approximately, from 0 to 0.1.} we have \textit{high-} and \textit{low-mirror} configurations. The \textit{standard} configuration (in which all planar coil currents are set to zero and the non-planar coil currents are set to the same value) and the high-mirror configuration feature particularly reduced levels of neoclassical transport and bootstrap current, respectively. These facts, in part, explain why they are the configurations most commonly employed for experiments and simulations (for a detailed description of the physical and technical features of W7-X configurations see \citet{Geiger2015}). In this work, we will use a high-mirror configuration. For future electromagnetic extensions of this benchmark and because W7-X aims at operating at high $\left<\beta\right>$ \citep{Klinger2019}, we have considered a high-mirror equilibrium with $\left<\beta\right>\approx 3\%$. Specifically, the equilibrium is based on a fixed-boundary VMEC-calculation. This means that it does not use the coils definition file \texttt{mgrid} but employs a simplified high-mirror vacuum configuration boundary.
\begin{table}
\centering
\begin{tabular}{ c c c c c c c c c c c c c c c c c c c c c c c c}
    \toprule
  $a$ $[\mathrm{m}]$  & & & & $R_0$ $[\mathrm{m}]$  & & & &  $s_0$ & & & & {$B_{00}(s_0)$} $[\mathrm{T}]$ & & & &  $\iota(s_0)$ & & & & $-\frac{\sqrt{s_0}}{\iota(s_0)} \frac{d\iota}{d \sqrt{s}}(s_0)$ \\  \midrule
  0.494 & & & & 5.485 & & & & 0.64 & & & & 2.762 & & & & 0.910 & & & & -0.107  \\ \bottomrule
\end{tabular}
  \caption{Basic quantities of the magnetic configuration and the flux surface selected for the simulations. From left to right: effective minor radius, major radius, normalized toroidal magnetic flux, mode $B_{00}$ of the Fourier expansion of the magnetic field strength, rotational transform and global magnetic shear.}
  \label{table_surf}
\end{table}
\begin{figure}
    \centering
        \includegraphics[width=0.5\linewidth]{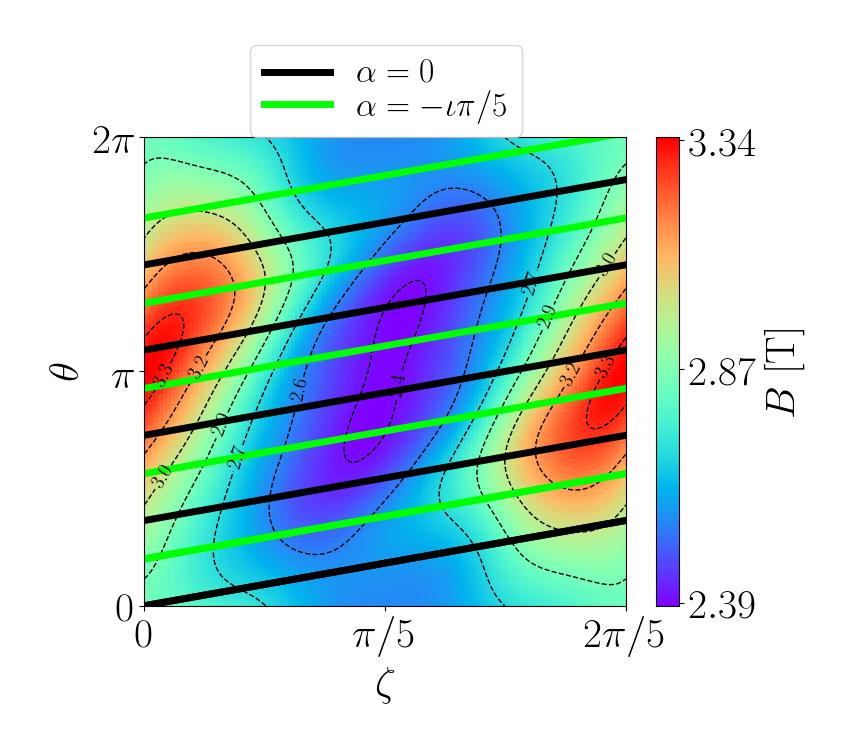}
   \caption{Schematic view of the magnetic field lines $\alpha=0$ (solid black line) and $\alpha=-\iota\pi/5$ (solid green line) extended along the five field periods of W7-X. The magnetic field strength is represented in the background.}
    \label{fig.lines}
\end{figure}
\begin{figure}
    \centering 
    \begin{subfigure}[b]{0.85\linewidth}
        \centering
        \includegraphics[width=0.9\linewidth]{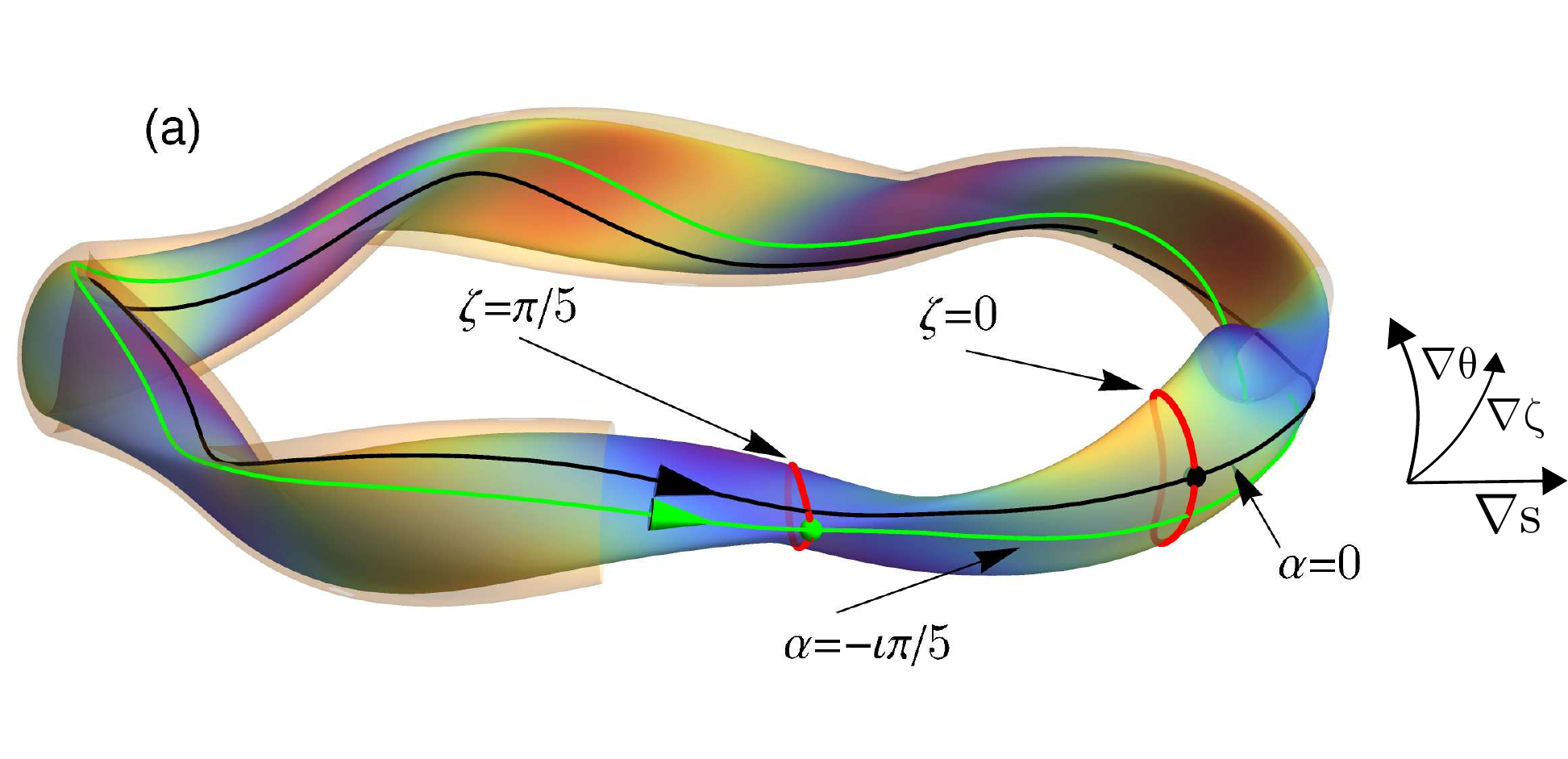}        
    \end{subfigure}
        \begin{subfigure}[b]{0.1\linewidth}
        \centering
        \includegraphics[width=\linewidth]{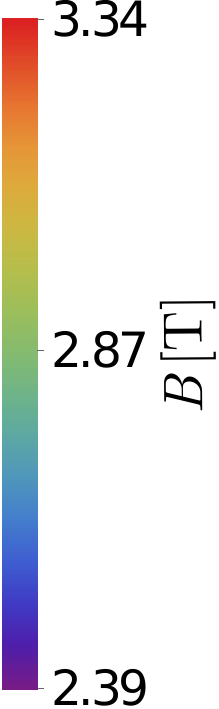}       
    \end{subfigure}
    \begin{subfigure}[b]{0.45\linewidth}        
        \centering 
        \includegraphics[width=\linewidth]{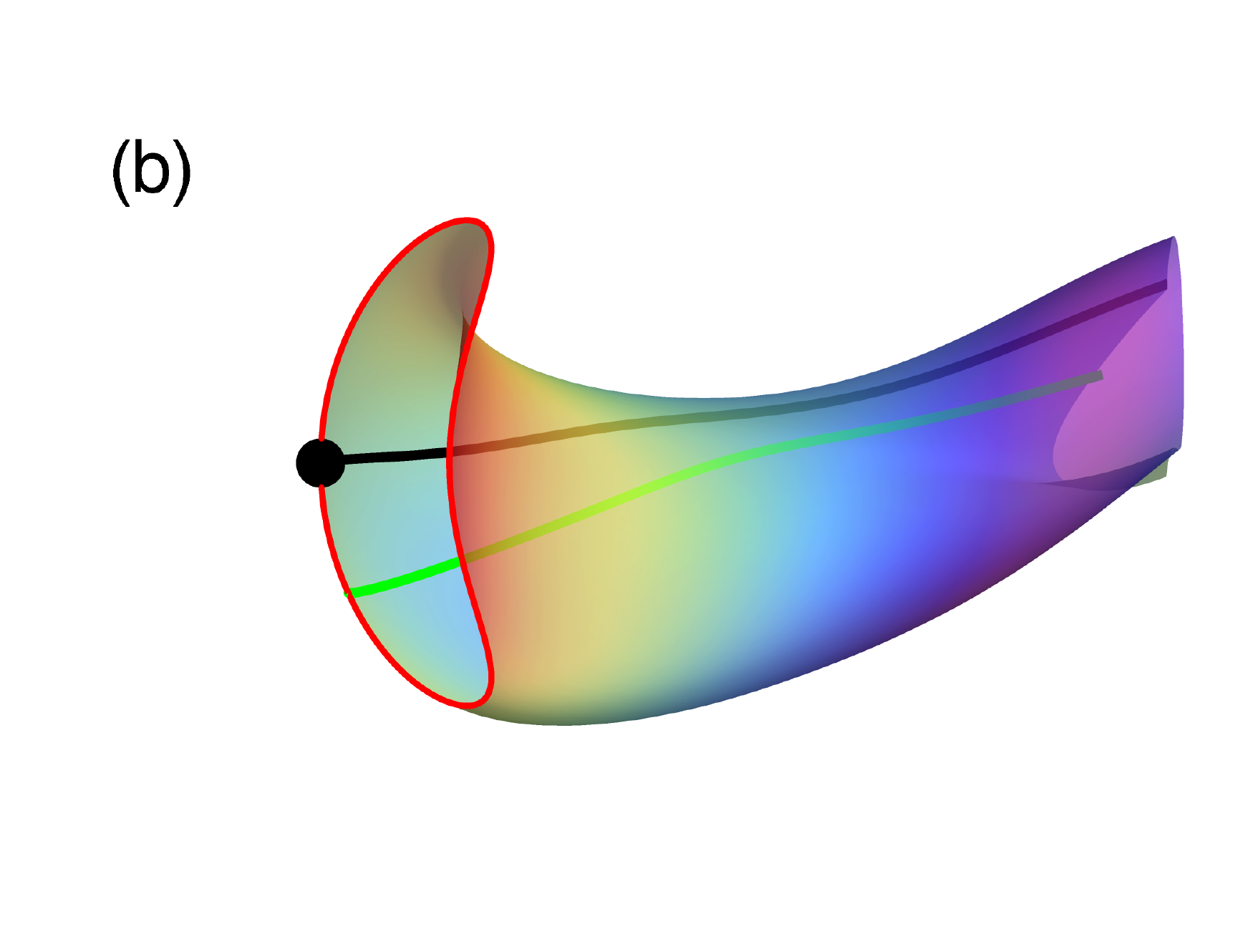}
    \end{subfigure}
    \begin{subfigure}[b]{0.45\linewidth}      
        \centering 
        \includegraphics[width=1.3\linewidth]{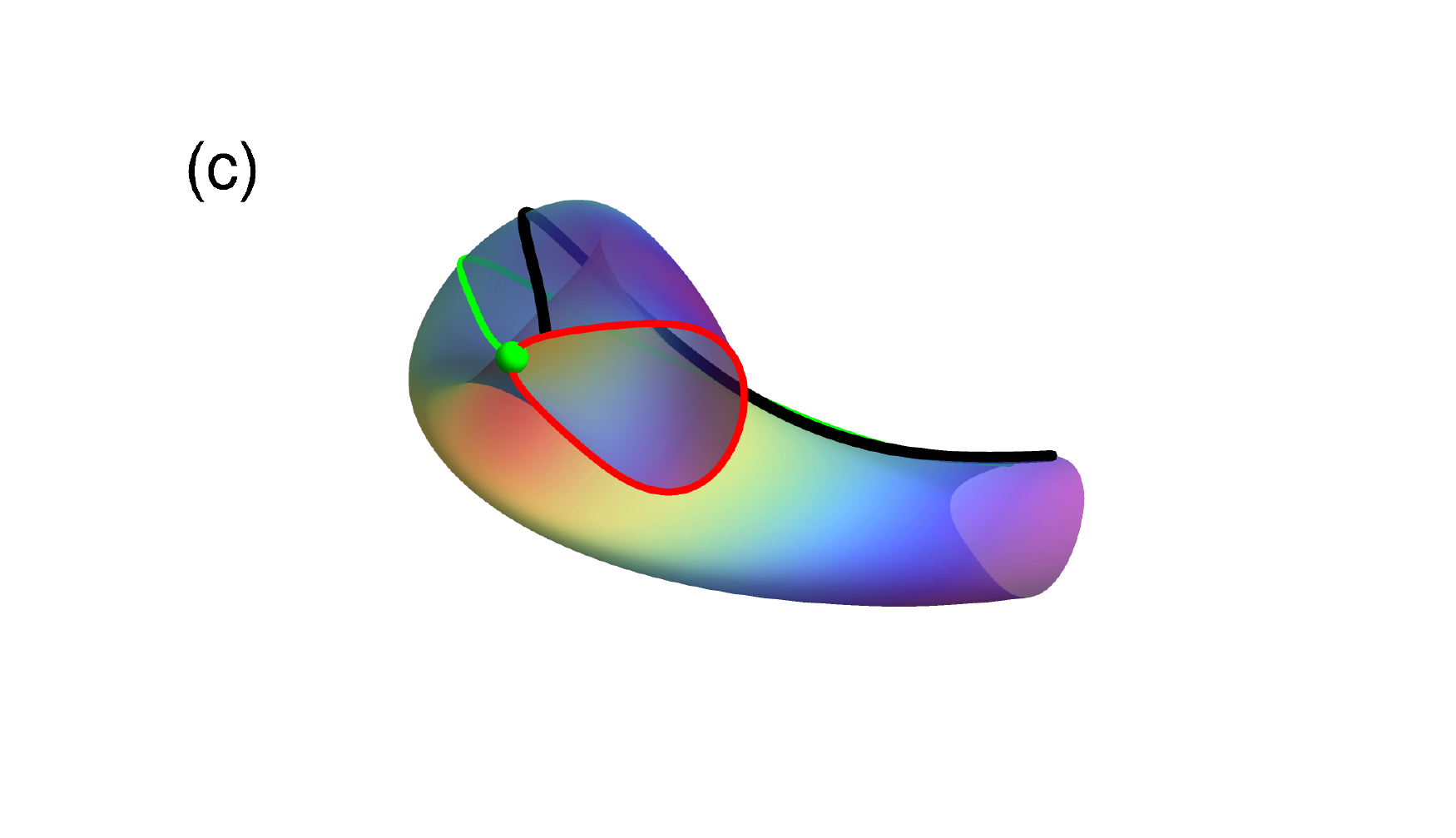}
    \end{subfigure}
    \caption{3D view of the surface $s_0=0.64$ (a), together with the field line $\alpha=0$ (solid black line), the field line $\alpha=-\iota\pi/5$ (solid green line) and the last closed flux surface $s=1$ as a semi-transparent halo. Details of two toroidal cuts of the flux surface $s_0=0.64$ are also given, showing a bean-shaped section (b) and a triangular section (c).}
    \label{fig.tubes}
\end{figure}
Some basic quantities of this magnetic configuration and the flux surface selected for our simulations are listed in table \ref{table_surf}. In Appendix \ref{aped.A}, the input parameters necessary to produce this {fixed-boundary equilibrium} with VMEC are provided.

\begin{figure}
    \centering
        \includegraphics[width=0.9\linewidth]{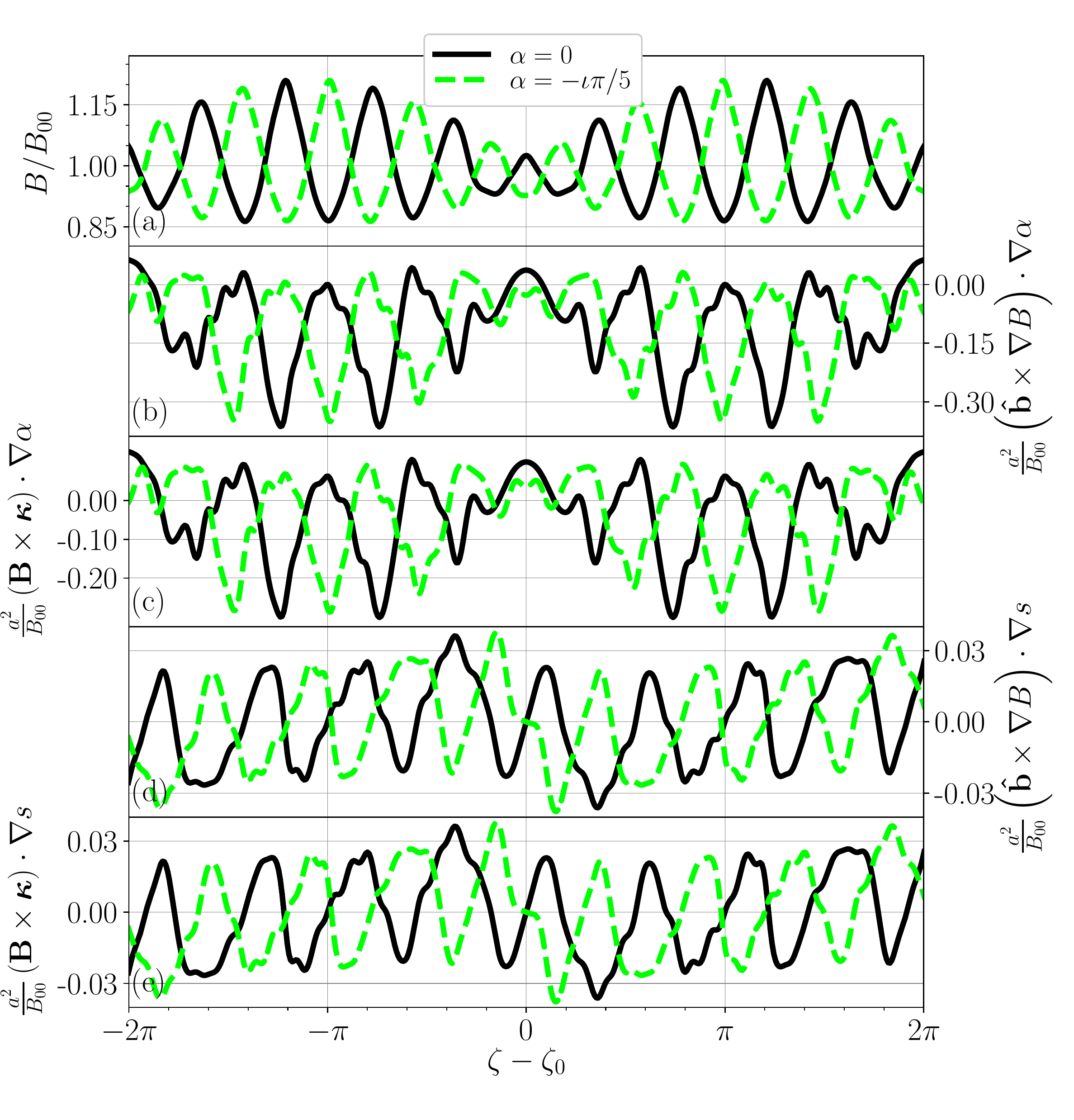}        
   \caption{Normalized geometric quantities in the range $\zeta-\zeta_0=[-2\pi,2\pi]$ for the surface $s_0=0.64$ along the field lines $\alpha=0$ (solid black line) and $\alpha=-\iota\pi/5$ (dashed green line). The magnetic field strength is represented in (a); the projections of $\mathbf{\hat{b}}\times\nabla B$ and $\mathbf{B}\times\pmb{\kappa}$ along the binormal direction are represented in (b) and (c), respectively; the projections of $\mathbf{\hat{b}}\times\nabla B$ and $\mathbf{B}\times\pmb{\kappa}$ along the radial direction are represented in (d) and (e), respectively.}
    \label{fig.geo}
\end{figure}

In stellarators, different magnetic field lines on a flux surface are not equivalent. In general, gyrokinetic simulations run on different flux tubes lead to different results. For this reason, this work includes simulations in two different flux tubes, the one that extends along the widely simulated field line $\alpha=0$ and the less common choice $\alpha=-\iota\pi/5$. The field line $\alpha=0$ is centered with respect to the so-called equatorial plane, $\theta=0$, and the bean-shaped toroidal plane $\zeta=0$, hence the name bean flux tube. The field line $\alpha=-\iota\pi/5$ is centered with respect to the equatorial plane and the triangular toroidal plane $\zeta=\pi/5$, hence the name triangular flux tube. A schematic view of these field lines for the flux surface $s_0=0.64$ is given in figure \ref{fig.lines}, where they are represented on a $(\theta,\zeta)$ plane with the magnetic field strength referred to the color scale. A 3D view of the surface $s_0=0.64$ is shown in figure \ref{fig.tubes} (a) together with a sketch of our choice of flux coordinates and the field lines $\alpha=0$ and $\alpha=-\iota\pi/5$. Figures \ref{fig.tubes} (b) and \ref{fig.tubes} (c) show some details of figure \ref{fig.tubes} (a). The geometric quantities required to solve equations (\ref{norm_gyro_eq}) and (\ref{quasi}) for the mentioned flux tubes are represented against the $\zeta$ coordinate, centered with respect to $\zeta_0=\zeta(\theta=0)$, in figures \ref{fig.geo} (a)-(e). In figure \ref{fig.geo} (a) it is observed that the magnetic field strength is symmetric with respect to $\zeta=\zeta_0$. This symmetry is also seen in the quantities $\left(\mathbf{\hat{b}}\times\nabla B\right)\cdot \nabla \alpha$ and $\left(\mathbf{B}\times\pmb{\kappa}\right)\cdot \nabla \alpha$, which are represented in figures \ref{fig.geo} (b) and \ref{fig.geo} (c), respectively. In figures \ref{fig.geo} (d) and \ref{fig.geo} (e) we see that $\left(\mathbf{B}\times\pmb{\kappa}\right)\cdot\nabla s= \left(\mathbf{\hat{b}}\times\nabla B\right)\cdot\nabla s$, as is the case for any ideal MHD equilibria. In what follows, we will refer to the direction of $\nabla\alpha$ as binormal direction and to the direction of $\nabla s$ as radial direction. The description of the flux tubes is complete with the specification of their length, which has been defined as the number of turns in the poloidal direction, $N_{\theta}$. This length is chosen to correctly resolve the electrostatic potential $\hat{\varphi}_{\mathbf{k}}$ along the flux tube. Since the localization of the electrostatic potential varies with the wavenumber, different values of $N_{\theta}$ have been considered in each test. 
\begin{table}
\centering
\begin{tabular}{c c c c c c c c c c c}
    \toprule
  & Flux tube & $\left[N_{\theta}^{m},N_{\theta}^{M}\right]$ & $a/L_{T_i}$ &  $a/L_{n_i}$ & $N_z$ & $N_{v_{\|}}$ & $N_{\mu}$ & $N_{\mathbf{k},j}$ & $\Delta tv_{\mathrm{{th,i}}}/a$ & Compared  \\ \bottomrule
  \multirow{2}{*}{Test 1.} &  \multirow{2}{*}{bean} & \multirow{2}{*}{[1 , 6]} & \multirow{2}{*}{3} & \multirow{2}{*}{1} & \multirow{2}{*}{256} & \multirow{2}{*}{36} & \multirow{2}{*}{24} & \multirow{2}{*}{Off} & \texttt{stella} 0.15 & $\gamma(k_x)$, $\omega(k_x)$ \\ &&&&&&&&& \texttt{GENE} 0.14 & $\gamma(k_y)$, $\omega(k_y)$ \\ \midrule
  \multirow{2}{*}{Test 2.} &  \multirow{2}{*}{triangular} & \multirow{2}{*}{[4 , 6]} & \multirow{2}{*}{3} & \multirow{2}{*}{1} & \multirow{2}{*}{512} & \multirow{2}{*}{36} & \multirow{2}{*}{24} & \multirow{2}{*}{Off} &  \texttt{stella} 0.15  & \multirow{2}{*}{$\gamma(k_x)$, $\omega(k_x)$} \\ &&&&&&&&& \texttt{GENE} 0.14\\ \midrule
  \multirow{2}{*}{Test 3.} &  \multirow{2}{*}{bean} & \multirow{2}{*}{[2 , 8]} & \multirow{2}{*}{0} & \multirow{2}{*}{3} & \multirow{2}{*}{512} & \multirow{2}{*}{36} & \multirow{2}{*}{24} & \multirow{2}{*}{Off} & \texttt{stella} 0.04 & $\gamma(k_y)$, $\omega(k_y)$\\ &&&&&&&&&  \texttt{GENE} 0.004 & $|\hat{\varphi}_{\mathbf{k}}|(z)$\\ \midrule    
  \multirow{2}{*}{Test 4.} &  \multirow{2}{*}{bean} & \multirow{2}{*}{[4 , 4]} & \multirow{2}{*}{0} & \multirow{2}{*}{0} & \multirow{2}{*}{512} & \multirow{2}{*}{256} & \multirow{2}{*}{32} & \multirow{2}{*}{Off} & \texttt{stella} 0.15 &  \multirow{2}{*}{$\langle\Re(\hat{\varphi}_{\mathbf{k}})\rangle_{z}(t)$} \\ &&&&&&&&& \texttt{GENE} 0.1\\ \midrule
  \multirow{3}{*}{Test 5.} &  \multirow{3}{*}{bean\footnote[5]{}} & \multirow{3}{*}{[1 , 1]} & \multirow{3}{*}{3} & \multirow{3}{*}{1} & \multirow{3}{*}{128} & \multirow{3}{*}{60} & \multirow{3}{*}{24} & \multirow{3}{*}{On} & \multirow{2}{*}{\texttt{stella} 0.09} & $Q_i(t)$  \\ &&&&&&&&&  \multirow{2}{*}{\texttt{GENE} 0.09} & $\sum_{k_y}Q_i(k_x,k_y)$ \\ &&&&&&&&&& $\sum_{k_x}Q_i(k_x,k_y)$ \\ \bottomrule 

\end{tabular}
  \caption{Set of parameters used in each test. From left to right: flux tube, minimum and maximum number of $N_{\theta}$, normalized ion temperature and density gradients, number of divisions in the grid of $z$, $v_{||}$ and $\mu$, presence of non-linear term, time step size and quantities compared with both codes.} 
  \label{table_param}
\end{table}
The maximum and minimum values of $N_{\theta}$, $N_{\theta}^M$ and $N_{\theta}^m$, respectively, needed in each test are indicated in table \ref{table_param}, together with other parameters that define the simulations. These include the normalized ion temperature and density gradients\footnote[1]{The normalized electron temperature and density gradients are set to zero, {$a/L_{T_e}=a/L_{n_e}=0$, for every test with adiabatic electrons, which excludes test 3}. The values given to these quantities in test 3 will be specified in section \ref{sec.lineal}.} \footnotetext[5]{Nonlinear results obtained with \texttt{stella} in the triangular flux tube are also included in section 5.}$\left(a/L_{T_i}=-\dd \ln T_i/\dd \sqrt{s}\right.$ and $\left.a/L_{n_i}=-\dd\ln {n_i}/\dd\sqrt{s}\right)$, the number of divisions in the parallel coordinate grid $(N_z)$, the number of divisions in the velocity grid $(N_{v_{||}},N_{\mu}$), the presence of nonlinear term ($N_{\mathbf{k},j}$), the time step size used for the calculation of the most unstable mode in each simulation and the different quantities compared in each test, where $\gamma$ refers to the growth rate, $\omega$ refers to the real frequency, $\langle .\rangle_{z}$ means a line average and $Q_i$ is the ion heat flux. It is important to remark that, {in all simulations in this paper,} $Z_i=1$ and $T_i/T_e=1$ have been assumed. In what follows, the results of the tests are expressed in the coordinates used by \texttt{stella}, $\{x,y,z\}=\{a\sqrt{s},a\sqrt{s_0}\alpha,\zeta\}$.

%% file: files/Linear.tex
\section{Linear simulations}\label{sec.lineal} 

To perform a linear flux tube gyrokinetic simulation, both codes solve the system of equations consisting of (\ref{norm_gyro_eq}) and (\ref{quasi}) dropping the nonlinear term of (\ref{norm_gyro_eq}) and assuming $\hat{g}_\mathbf{k}$ and $\hat{\varphi}_{\mathbf{k}}$ to be proportional to $\exp\left[(\gamma -\mathrm{i} \omega)t\right]$, where $\gamma$ and $\omega$ are the growth rate and real frequency of each mode. In this section, four different linear tests are presented. In tests 1 and {2, ITGs are simulated. In test 3, density-gradient-driven TEMs are computed. Finally,} the simulations of test 4 include the collisionless relaxation of a zonal electrostatic potential. 

The linear properties of ITGs and TEMs in W7-X have been reported in a large number of studies by means of linear gyrokinetic simulations. \citet{Kornilov2004} studied the ITG structure and its stability with the global particle-in-cell code \texttt{EUTERPE}, \citet{Xanthopoulos2007_linear} and \citet{Proll2013} used the code \texttt{GENE} to study the effect of changes in the density gradient and temperature ratio on ITGs and TEMs and to look at the stability properties of W7-X, comparing with other devices. \texttt{GENE} has also been used in \citet{Proll2015} to investigate how stellarators can be optimized with respect to TEMs and in \citet{Alcusn2020} to analyze the growth rate of the instability as a function of the temperature and density scales for different configurations of W7-X. A theoretical study about the effects of these instabilities in non-axisymmetric devices and, particularly, in W7-X is summarized in \citet{Helander2015}.

\subsection{Test 1. Linear ITG simulations in the bean flux tube}

\begin{figure}
    \centering
        \includegraphics[width=0.48\linewidth]{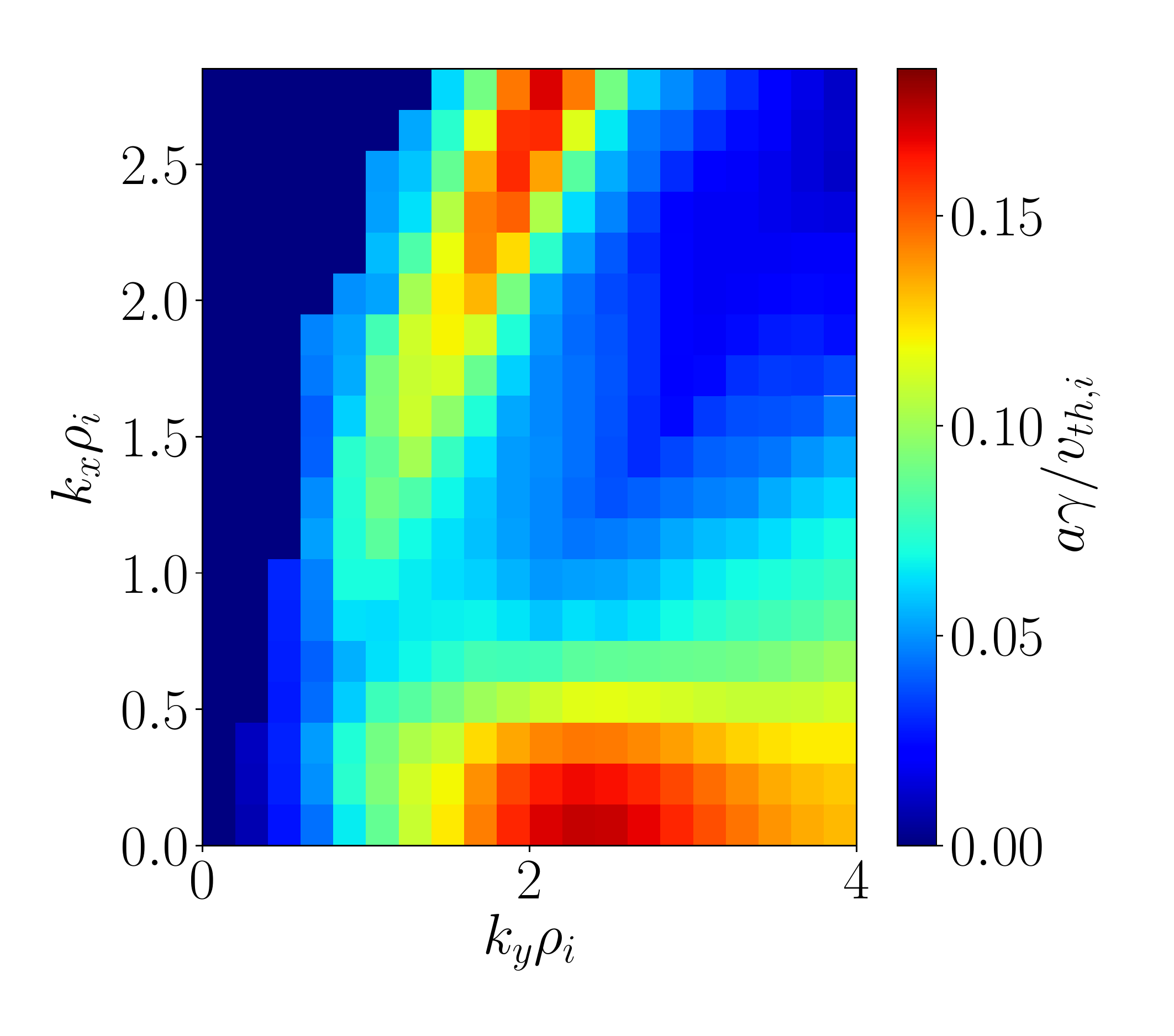}
    \caption{ITG stability map corresponding to test 1. It shows the growth rate computed with the code \texttt{stella} in the bean flux tube as a function of $k_x$ and $k_y$.}
    \label{fig.map_ITG_1}
\end{figure}
In this test, a linear ITG driven by a normalized ion temperature gradient $a/L_{T_i}=3$ with a normalized ion density gradient $a/L_{n_i}=1$ and adiabatic electrons is simulated in the bean flux tube (see table \ref{table_param}). In order to find the most unstable mode, a map containing the growth rate values as a function of the radial and binormal wavenumbers has been produced with \texttt{stella} and shown in figure \ref{fig.map_ITG_1}. In this map, two regions of large growth rate can be observed. While $N_{\theta}=1$ is enough to simulate the region with $k_x\rho_i\lesssim0.5$, $N_{\theta}=6$\footnote[1]{These lengths are the required ones if the flux tube is centered at $z=0$.} is required to simulate the one including $k_x\rho_i\gtrsim2$ due to the displacement in the $z$ direction of the parallel structure of the modes. The maximum growth rate found in this map is localized at $k_y\rho_i=2.1$. The codes \texttt{GENE} and \texttt{stella} have been used to compare the spectrum along $k_y$ for fixed $k_x$ and vice-versa, capturing this wavenumber in both scans.

\begin{figure}
    \centering
    \begin{subfigure}[b]{0.48\linewidth}        
        \centering
        \includegraphics[width=\linewidth]{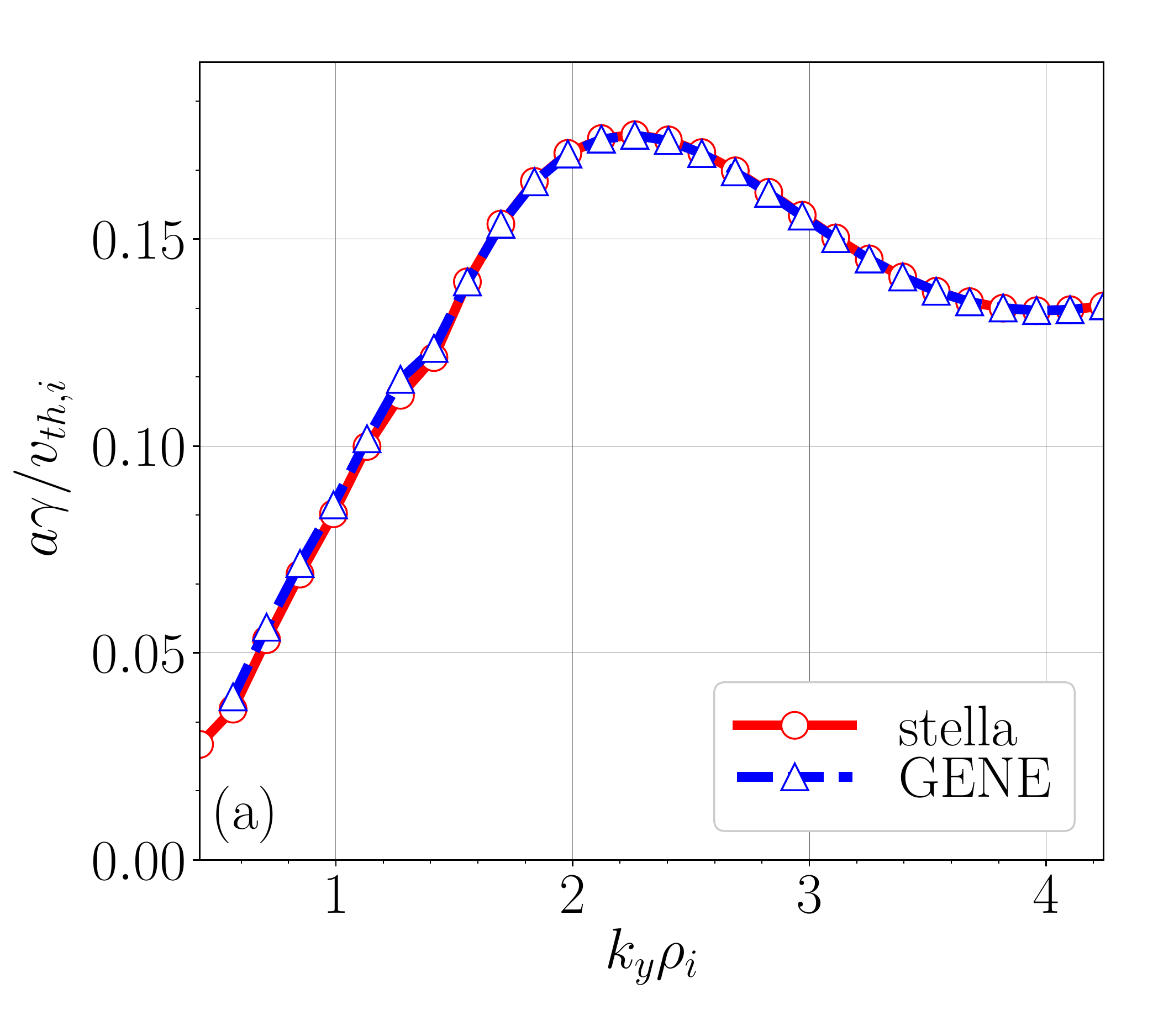}
    \end{subfigure}
    \begin{subfigure}[b]{0.48\linewidth}        
        \centering
        \includegraphics[width=\linewidth]{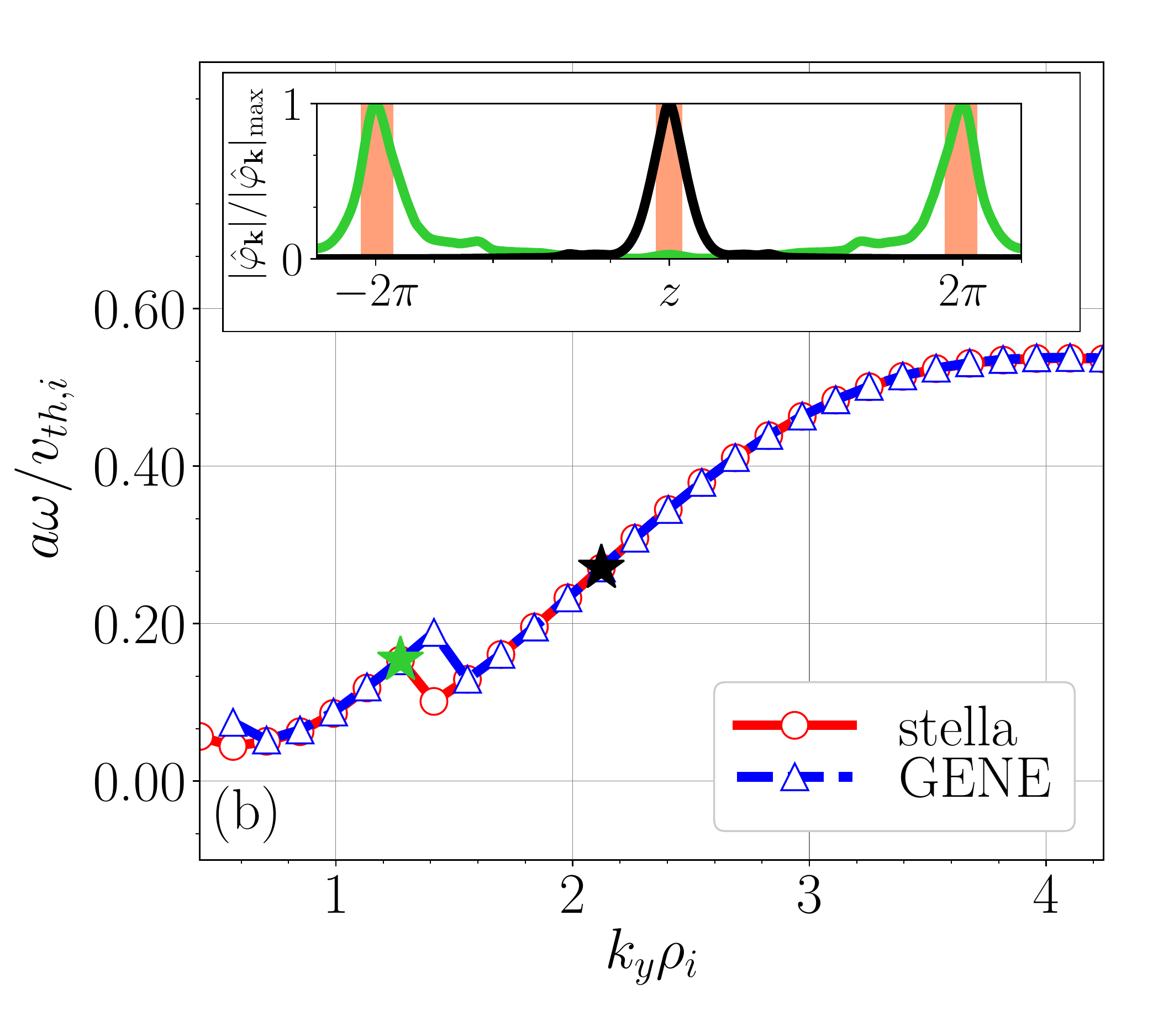}
    \end{subfigure}
    \caption{Linear growth rate (a) and real frequency (b) as a function of $k_y$ obtained for the ITG scenario studied in test 1 using \texttt{stella} (open circles linked by a solid red line) and \texttt{GENE} (open triangles linked by a dashed blue line) in the bean flux tube. The inset of figure (b) shows the structure of the modes $(k_x\rho_i,k_y\rho_i)=(0,1.3)$ (green line) and $(k_x\rho_i, k_y\rho_i)=(0,2.1)$ (black line) together with some bad curvature regions (shaded in red).}
    \label{fig.ITG.1}
\end{figure}

The comparison of growth rates and real frequencies as a function of $k_y$ for fixed $k_x=0$ is given in figures \ref{fig.ITG.1} (a) and \ref{fig.ITG.1} (b), respectively. These figures show an excellent agreement between \texttt{stella} and \texttt{GENE}. In figure \ref{fig.ITG.1} (b) it is seen that the frequency is positive for every simulated mode. ITG-driven modes are expected to propagate in the ion diamagnetic direction, i.e. $\omega\omega_{*,i}>0$, where $\omega_{*,i}$ is the ion diamagnetic frequency, defined as
\begin{equation}\label{eq.ion_diam}
    \omega_{*,i}=\frac{T_ik_y}{Z_ie\psi'_t}\frac{n'_i}{n_i}.
\end{equation}
As only positive values of $k_y$ are explored in this scan, $n'_i<0$ (see table \ref{table_param}) and $\psi'_t<0$ for the selected configuration (see the direction of $\mathbf{B}$ and the left-handed system sketched in figure \ref{fig.tubes} (a)), $\omega_{*,i}>0$. This proves that, indeed, in this test $\omega\omega_{*,i}>0$, thus the studied ITG-driven modes propagate in the ion diamagnetic direction. A closer look at figure \ref{fig.ITG.1} (b) shows a discontinuity in the frequency, which is associated with a change in the mode structure, defining two different branches of the ITG instability. This can be observed in the inset of figure \ref{fig.ITG.1} (b), which represents, as a function of $z$, computations of \texttt{stella} for the the modulus of the electrostatic potential normalized to its maximum value $\left(|\hat{\varphi}_{\mathbf{k}}|/|\hat{\varphi}_{\mathbf{k}}|_{\mathrm{max}}\right)$  for the modes with $k_y\rho_i=\{1.3,2.1\}$. These modes are strongly localized in the highlighted red bands, which correspond to \textit{bad curvature} regions, defined as those where
\begin{equation}
    \frac{k_yT'_i}{\psi'_t}\left(\mathbf{\hat{b}}\times\nabla B\right)\cdot\mathbf{k}_{\perp}>0.
\end{equation} 
If $k_x=0$ and $k_y>0$ these regions are the ones {where the quantity shown in figure \ref{fig.geo} (b) takes positive values}.
\begin{figure}
    \centering
    \begin{subfigure}[b]{0.48\linewidth}        
        \centering
        \includegraphics[width=\linewidth]{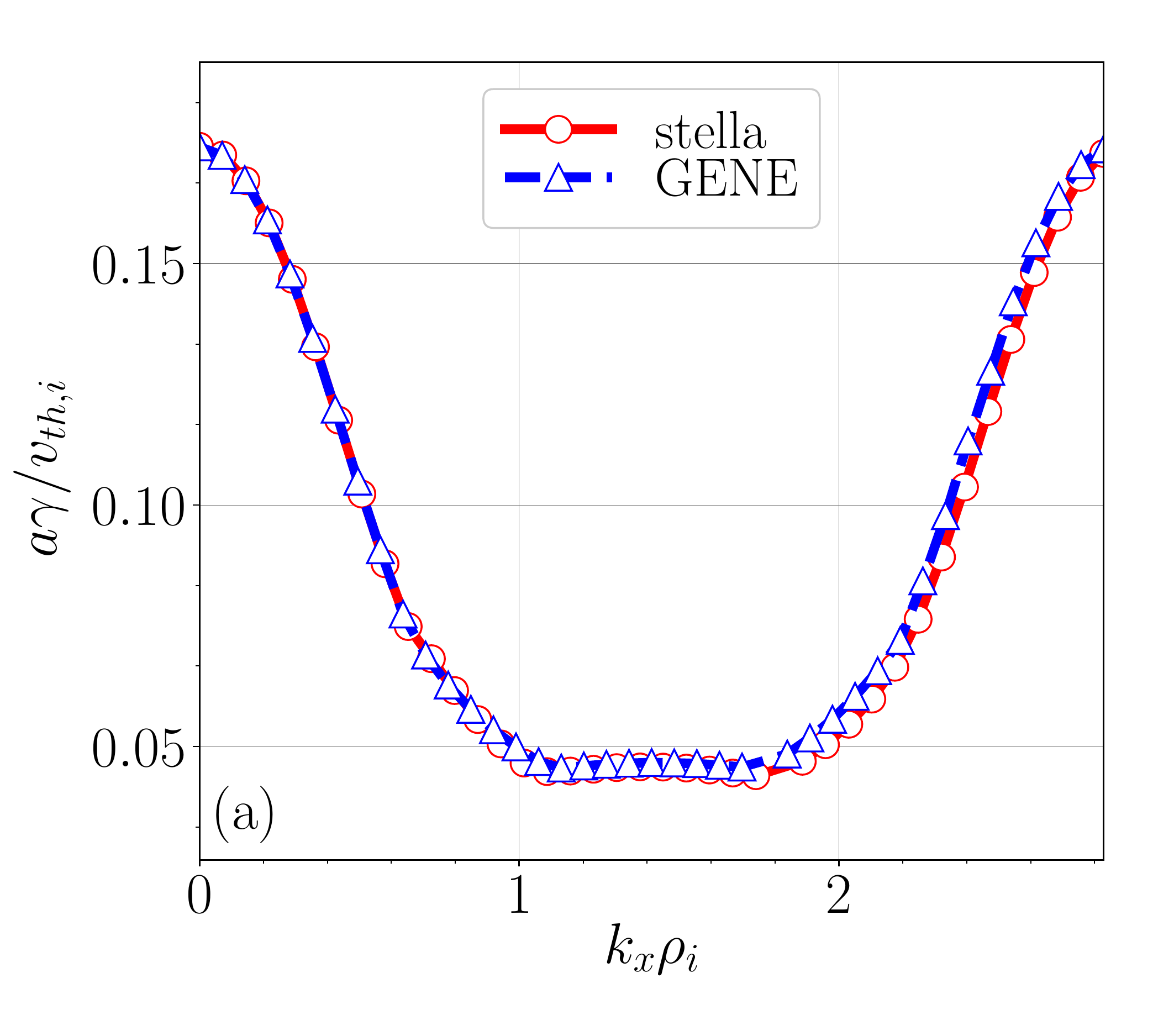}
    \end{subfigure}
    \begin{subfigure}[b]{0.48\linewidth}        
        \centering
        \includegraphics[width=\linewidth]{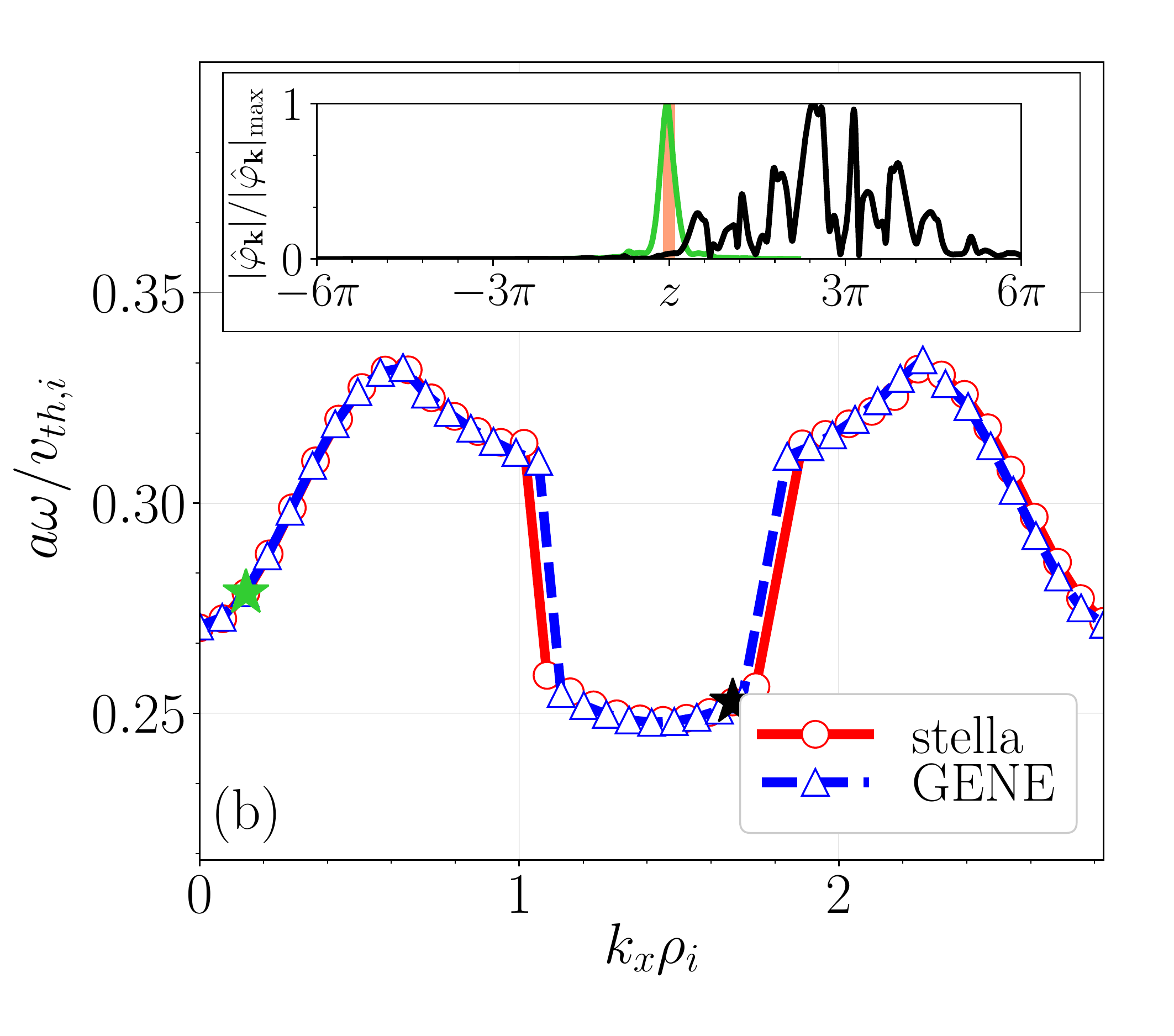}
    \end{subfigure}
    \caption{Linear growth rate (a) and real frequency (b) as a function of $k_x$ obtained for the ITG scenario studied in test 1 using \texttt{stella} (open circles linked by a solid red line) and \texttt{GENE} (open triangles linked by a dashed blue line) in the bean flux tube. The inset of figure (b) shows the structure of the modes $(k_x\rho_i,k_y\rho_i)=(0.2,2.1)$ (green line) and $(k_x\rho_i, k_y\rho_i)=(1.7,2.1)$ (black line), together with a bad curvature region (shaded in red).}
    \label{fig.ITG.1.5}
\end{figure}
The growth rates and real frequencies as a function of $k_x$, keeping $k_y\rho_i=2.1$, can be seen in figures \ref{fig.ITG.1.5} (a) and \ref{fig.ITG.1.5} (b), respectively. As in the $k_y$-spectra, every mode studied in this scan propagates in the ion diamagnetic direction, as it can be observed in figure \ref{fig.ITG.1.5} (b). This figure also shows a discontinuity in the frequency, giving rise to three different branches, located at $k_x\rho_i\in(0,1.0]$, $k_x\rho_i\in(1.0,1.8)$ and $k_x\rho_i\in[1.8,2.7)$. In the inset of figure \ref{fig.ITG.1.5} (b), the structure of the modes with $k_x\rho_i=\{0.2,1.7\}$ computed with \texttt{stella}, belonging to the first and central branches, respectively, are represented as a function of $z$. As observed in this inset, the electrostatic potential associated to the first branch is strongly localized and $N_{\theta}=1$ has been sufficient to capture the parallel structure of this mode. On the other hand, the electrostatic potential associated to the central branch spreads along $z$, making it necessary to extend the flux tube length up to $N_{\theta}=6$.


\subsection{Test 2. Linear ITG simulations in the triangular flux tube}

\begin{figure}
    \centering
    \includegraphics[width=0.48\linewidth]{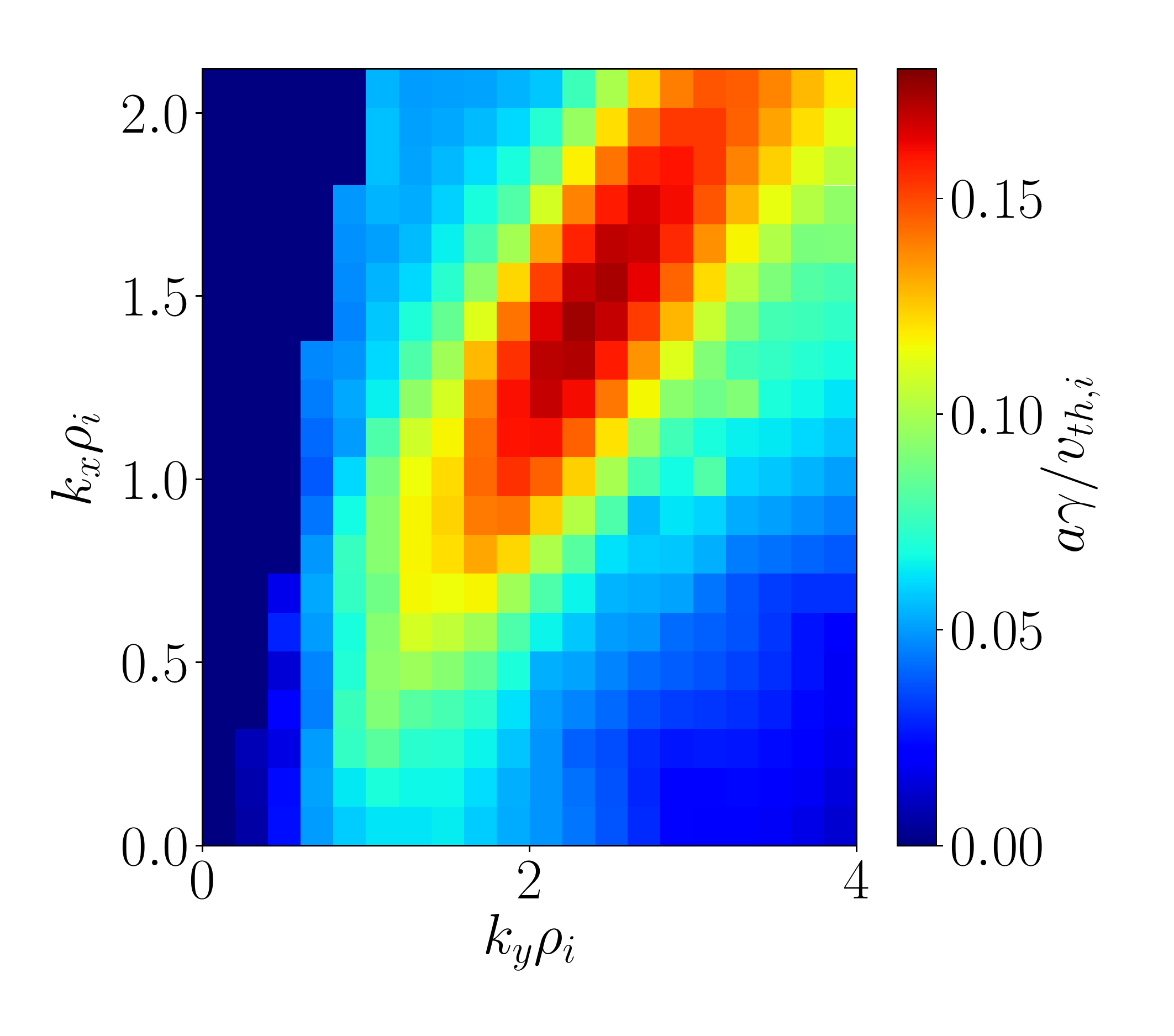}
    \caption{ITG stability map corresponding to test 2. It shows the growth rate computed with the code \texttt{stella} in the triangular flux tube as a function of $k_x$ and $k_y$.}
    \label{fig.map_ITG_2}
\end{figure}
In this test, a linear ITG instability driven by a normalized ion temperature gradient $a/L_{T_i}=3$ with a normalized ion density gradient $a/L_{n_i}=1$ and adiabatic electrons is simulated in the triangular flux tube (see table \ref{table_param}) .

As in the previous test, in order to find the most unstable mode, a map of the growth rate for each pair $(k_x,k_y)$ has been produced with \texttt{stella}, see figure \ref{fig.map_ITG_2}. As in the bean flux tube, the most unstable mode in this map is localized at $k_y\rho_i=2.1$. Interestingly, unlike in the bean flux tube, the maximum growth rate does not correspond to a mode with $k_x=0$. This figure also shows the triangular flux tube to be equally unstable as the bean one as $\gamma^{\mathrm{max}}_{\mathrm{bean}}/\gamma^{\mathrm{max}}_{\mathrm{triang}}\simeq 1$. The different localization in $k_x$ of the most unstable modes found in figures \ref{fig.map_ITG_1} and \ref{fig.map_ITG_2} implies that special care must be taken when comparing the linear stability properties of different flux tubes.
\begin{figure}
    \centering
    \begin{subfigure}[b]{0.48\linewidth}        
        \centering
        \includegraphics[width=\linewidth]{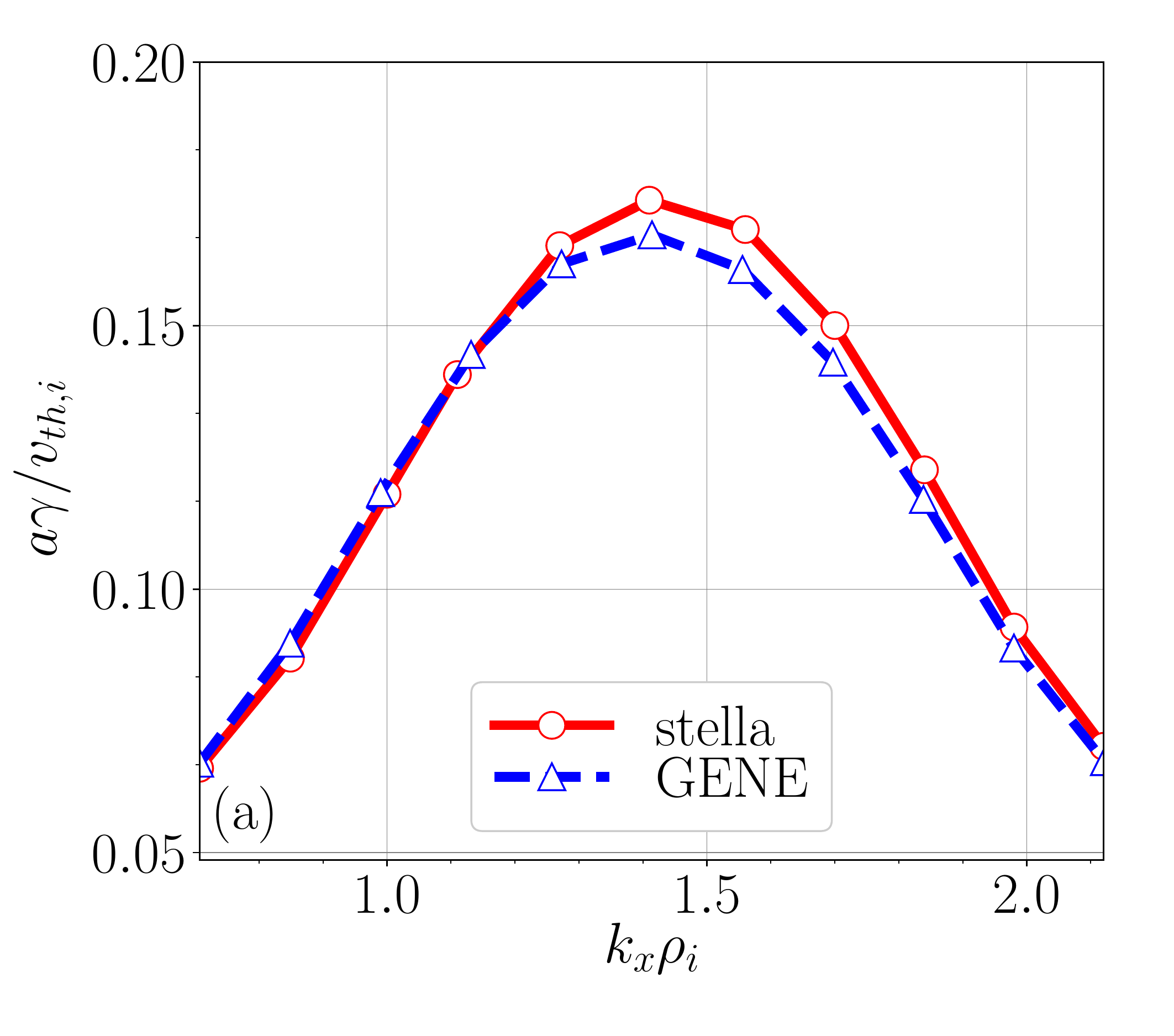}
    \end{subfigure}
    \begin{subfigure}[b]{0.48\linewidth}        
        \centering
        \includegraphics[width=\linewidth]{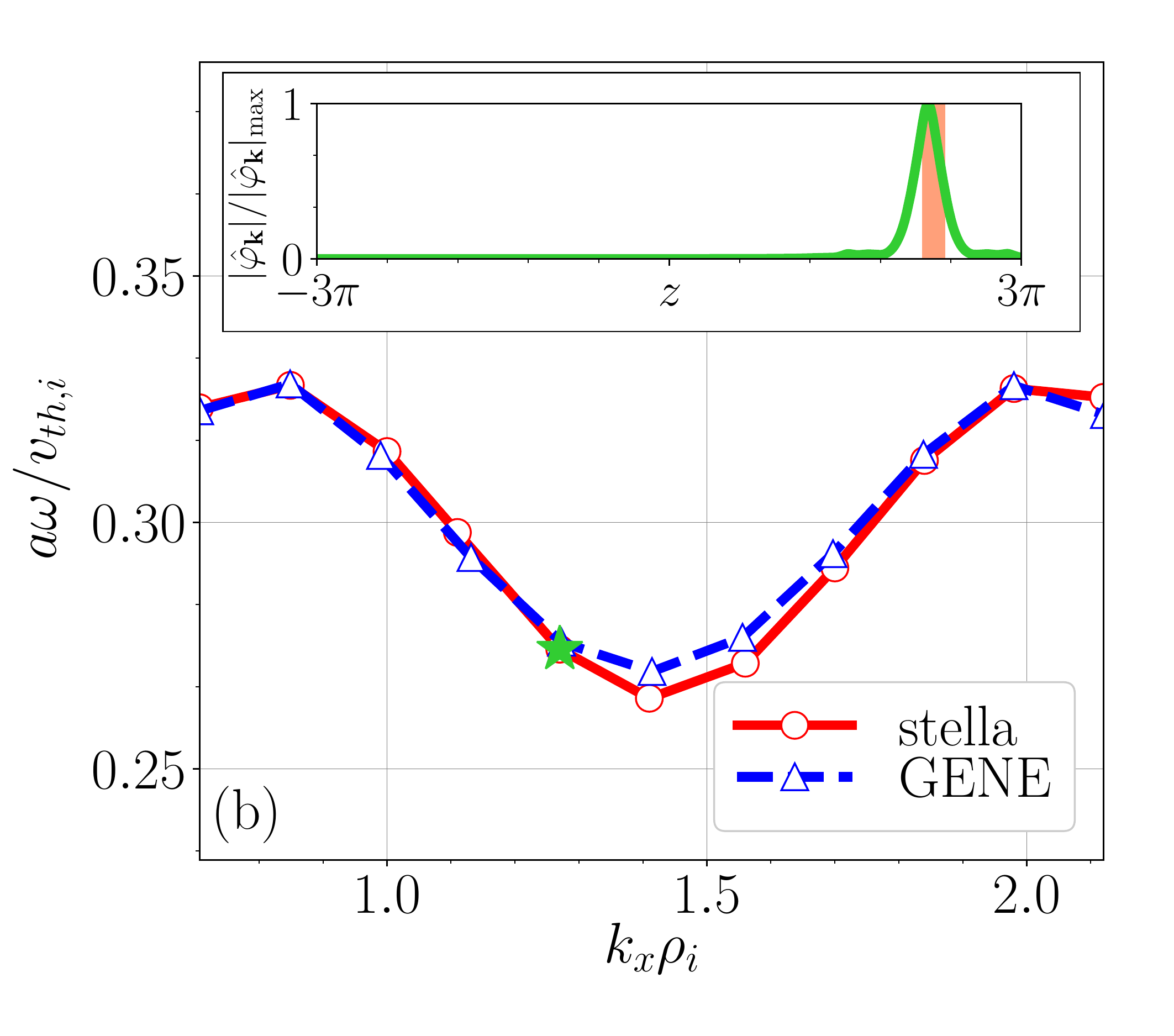}
    \end{subfigure}
    \caption{Linear growth rate (a) and real frequency (b) as a function of $k_x$ obtained for the ITG scenario studied in test 2 using \texttt{stella} (open circles linked by a solid red line) and \texttt{GENE} (open triangles linked by a dashed blue line) in the triangular flux tube. The inset of figure (b) shows the structure of the mode $(k_x\rho_i,k_y\rho_i)=(1.2,2.1)$ (green line) together with a bad curvature region (shaded in red).}
    \label{fig.ITG.2}
\end{figure}

A scan along $k_x$, keeping $k_y\rho_i=2.1$, has been performed with both codes, representing the growth rates and real frequencies in figures \ref{fig.ITG.2} (a) and \ref{fig.ITG.2} (b), respectively. Although not as close as in the bean flux tube, the agreement between \texttt{stella} and \texttt{GENE} is still remarkable. As in the previous test, these modes propagate in the ion diamagnetic direction. In this scan, the localization of the electrostatic potential moves to higher values of $z$ when increasing $k_x$, making it necessary to extend the flux tube length up to $N_{\theta}=6$. The inset of figure \ref{fig.ITG.2} (b) includes the parallel structure of the mode with $k_x\rho_i=1.2$ obtained with \texttt{stella}, together with the bad curvature region where this mode is localized.

\begin{figure}
    \centering
    \begin{subfigure}[b]{0.48\linewidth}        
        \centering
        \includegraphics[width=\linewidth]{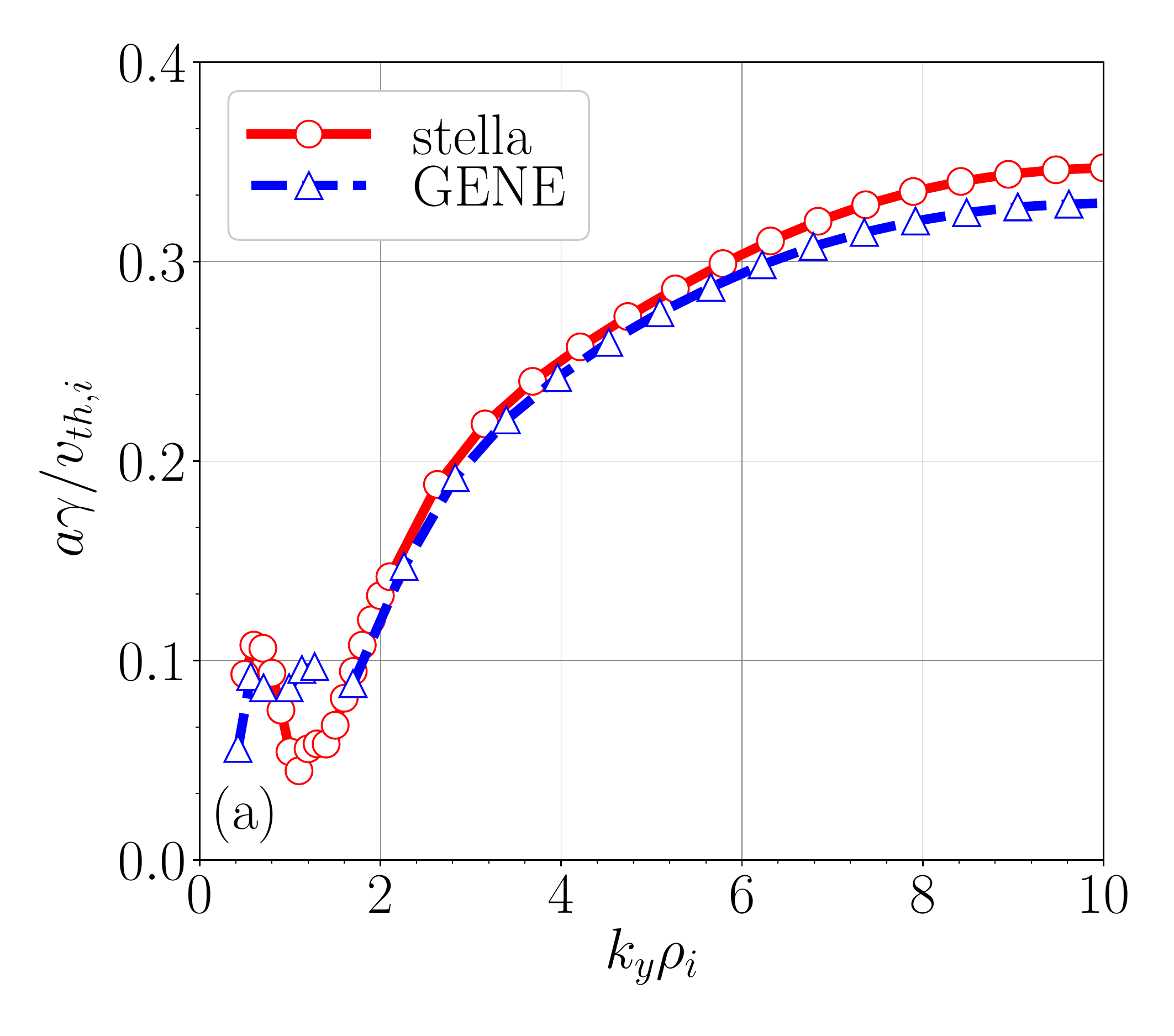}
    \end{subfigure}
    \begin{subfigure}[b]{0.48\linewidth}        
        \centering
        \includegraphics[width=\linewidth]{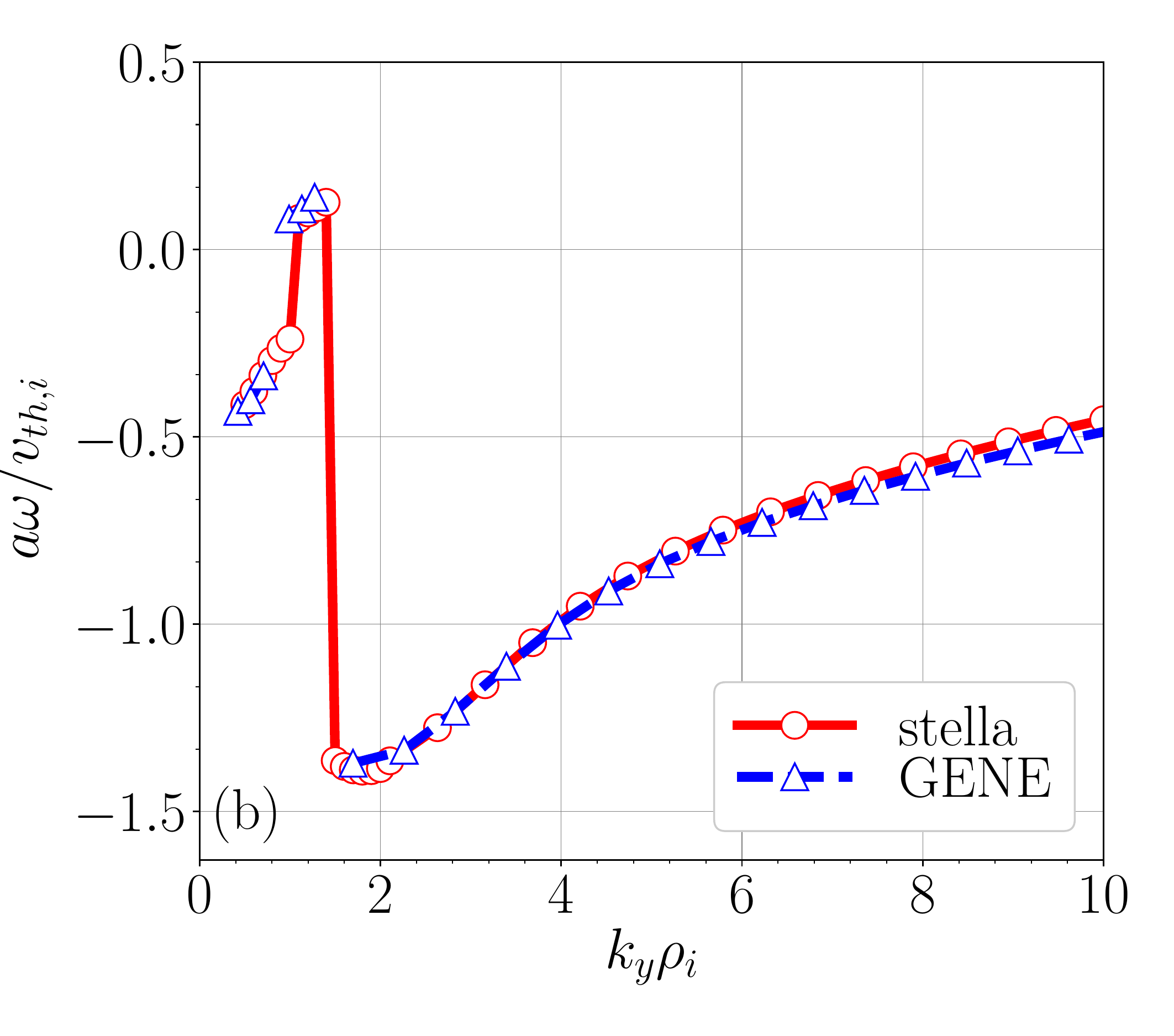}
    \end{subfigure} 
    \caption{Linear growth rate (a) and real frequency (b) as a function of $k_y$ obtained for the {instabilities} studied in test 3 using \texttt{stella} (open circles linked by a solid red line) and \texttt{GENE} (open triangles linked by a dashed blue line) in the bean flux tube.}
    \label{fig.TEM}
\end{figure}

\begin{figure}
    \centering
    \includegraphics[width=\linewidth]{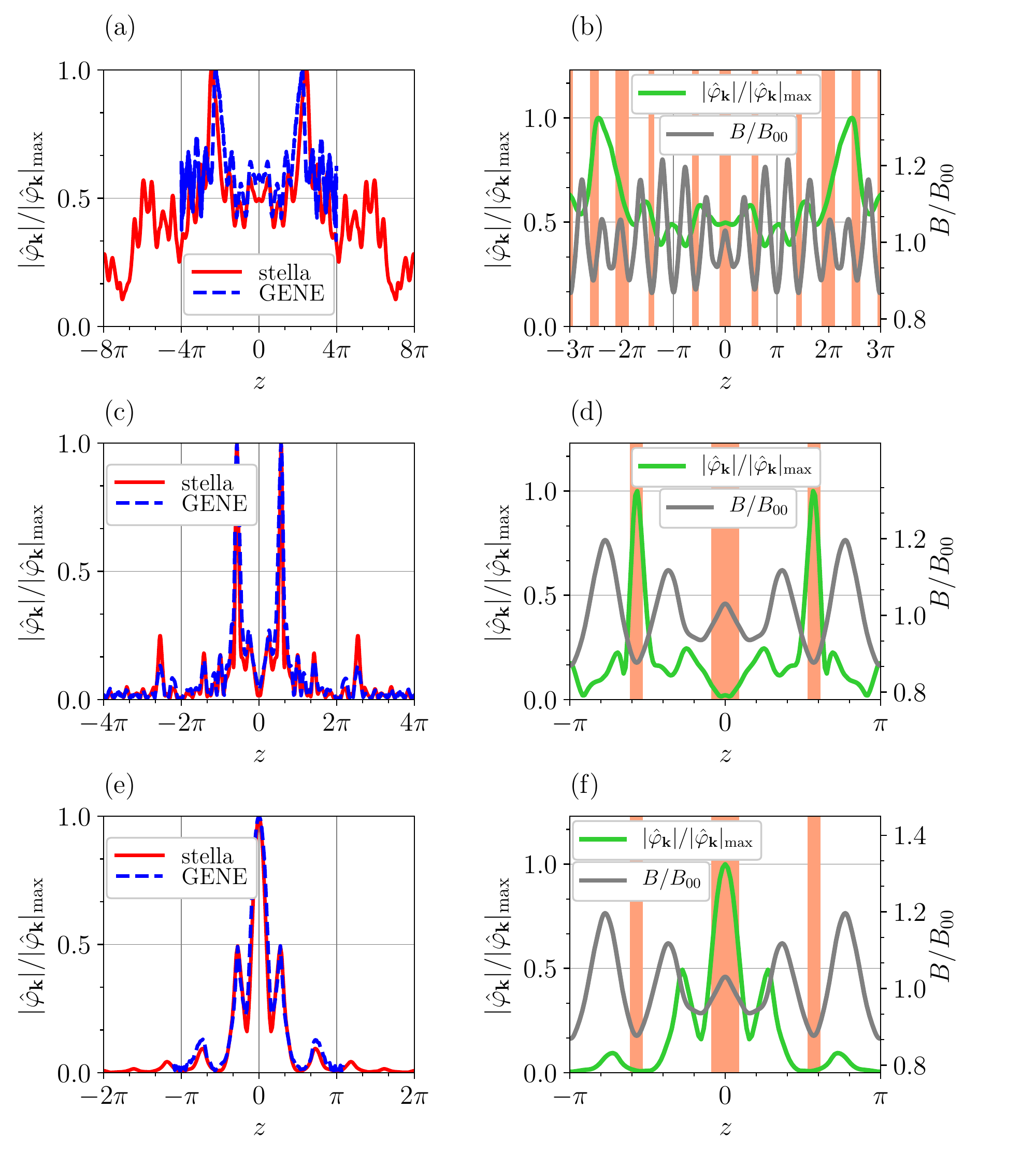}
    \caption{{Normalized modulus of the electrostatic potential computed with \texttt{stella} (solid red line) and \texttt{GENE} (dashed blue line) as a function of $z$ over the entire length of the flux tube for some modes simulated in test 3; specifically, we are representing the modes $(k_x\rho_i, k_y\rho_i)= (0,0.4)$ (a), $(0,1.3)$ (c) and $(0,4.7)$ (e).} The structures calculated with \texttt{stella} are {shown} as solid green lines in narrower $z$ ranges in figures (b), (d) and (f), respectively, together with the normalized magnetic field strength (grey line) and the bad curvature regions (shaded in red).}
    \label{fig.potential_TEM}
\end{figure}

\subsection{Test 3. Linear {density-gradient-driven} TEM simulations in the bean flux tube}
In the third test, {kinetic electrons are included. We study linear instabilities driven by} normalized electron and ion density gradients $a/L_{n_e}=a/L_{n_i}=3$. In order to avoid the presence of temperature gradient driven modes, the electron and ion temperature gradients have been set to zero, $a/L_{T_e}=a/L_{T_i}=0$ (see table \ref{table_param}). {We will refer to modes studied in this subsection as density-gradient-driven TEMs.} It is worth noting (see $\Delta t v_{\mathrm{th},i}/a$ in table \ref{table_param}) how the mixed implicit-explicit method employed by \texttt{stella} allows a larger time step in these simulations than the explicit scheme used in \texttt{GENE}.

The growth rate and real frequency values as a function of $k_y$, keeping $k_x=0$, are shown in figures \ref{fig.TEM} (a) and \ref{fig.TEM} (b), respectively. As observed in these figures, there is a remarkable agreement between \texttt{stella} and \texttt{GENE} results {at moderate to large $k_y\rho_i$}. In figure \ref{fig.TEM} (b) it can be seen that {these modes} can propagate both in the electron\footnote{{Since $Z_i=1$ and $T_i/T_e=1$, we have $\omega_{*,e}=-\omega_{*,i}$.}} and ion diamagnetic directions, depending on the wavenumber.  A closer look at this figure allows to clearly distinguish three different branches. The electrostatic potentials associated to the modes with $k_y\rho_i=\{0.4,1.3,4.7\}$, belonging each one to a different branch, are represented as a function of $z$ for the whole length of the flux tube in figures \ref{fig.potential_TEM} (a), (c) and (e), obtaining a good agreement between both codes. The same structures found in figures \ref{fig.potential_TEM} (a), (c) and (e) are represented in figures \ref{fig.potential_TEM} (b), (d) and (f), respectively, in a narrower $z$ range, together with the normalized magnetic field strength and the bad curvature regions. The parallel structure of the modes belonging to the first branch, in the range $k_y\rho_i=(0,1.1]$, has a particular shape (figures \ref{fig.potential_TEM} (a) and \ref{fig.potential_TEM} (b)), which can be identified with some structures discussed in \citet{Prollthesis}. To resolve this electrostatic potential with \texttt{stella}, we have increased the flux tube length up to $N_{\theta}=8$. In order to study the second branch, in the narrow range of $k_y\rho_i=(1.1,1.4]$, the flux tube has been extended up to $N_{\theta}=4$ with both codes. Finally, $N_{\theta}=2$ has been sufficient for the study of the third {branch}, covering from $k_y\rho_i=1.4$ to the end of the simulated range.


\subsection{Test 4. Zonal-flow relaxation in the bean flux tube}

Finally, we address the so-called Rosenbluth-Hinton test \citep{Rosenbluth1998}, which consists in the study of the linear collisionless time evolution of the zonal components of the potential, i.e. those with $k_y=0$, from their value at the initial time $t=0$ to {their value when $t\rightarrow \infty$. The theoretical study} of the zonal flow response in stellarators has been addressed in \citet{Sugama2005}; \citet{Mishchenko2008}; \citet{Helander2011}; \citet{Monreal2016}; \citet{Monreal2017} and \citet{Smoniewski2021}. In non-axisymmetric devices, the relaxation of a zonal potential perturbation typically shows a damped oscillation, reaching a stationary residual level at $t\rightarrow\infty$. The damped oscillation involves two different frequencies with different time scales: the geodesic acoustic mode (GAM) oscillation (which is also found in tokamaks) and a low frequency oscillation characteristic of the non-axisymmetric geometry of the stellarator.  This characteristic low frequency oscillation of the time evolution of the potential, only predicted for $k_x\rho_i\ll 1$ (\citet{Mishchenko2008}; \citet{Helander2011}; \citet{Monreal2017}), has been experimentally identified in the TJ-II stellarator in \citet{Alonso2017}. 

\begin{figure}
    \centering
    \includegraphics[width=\linewidth]{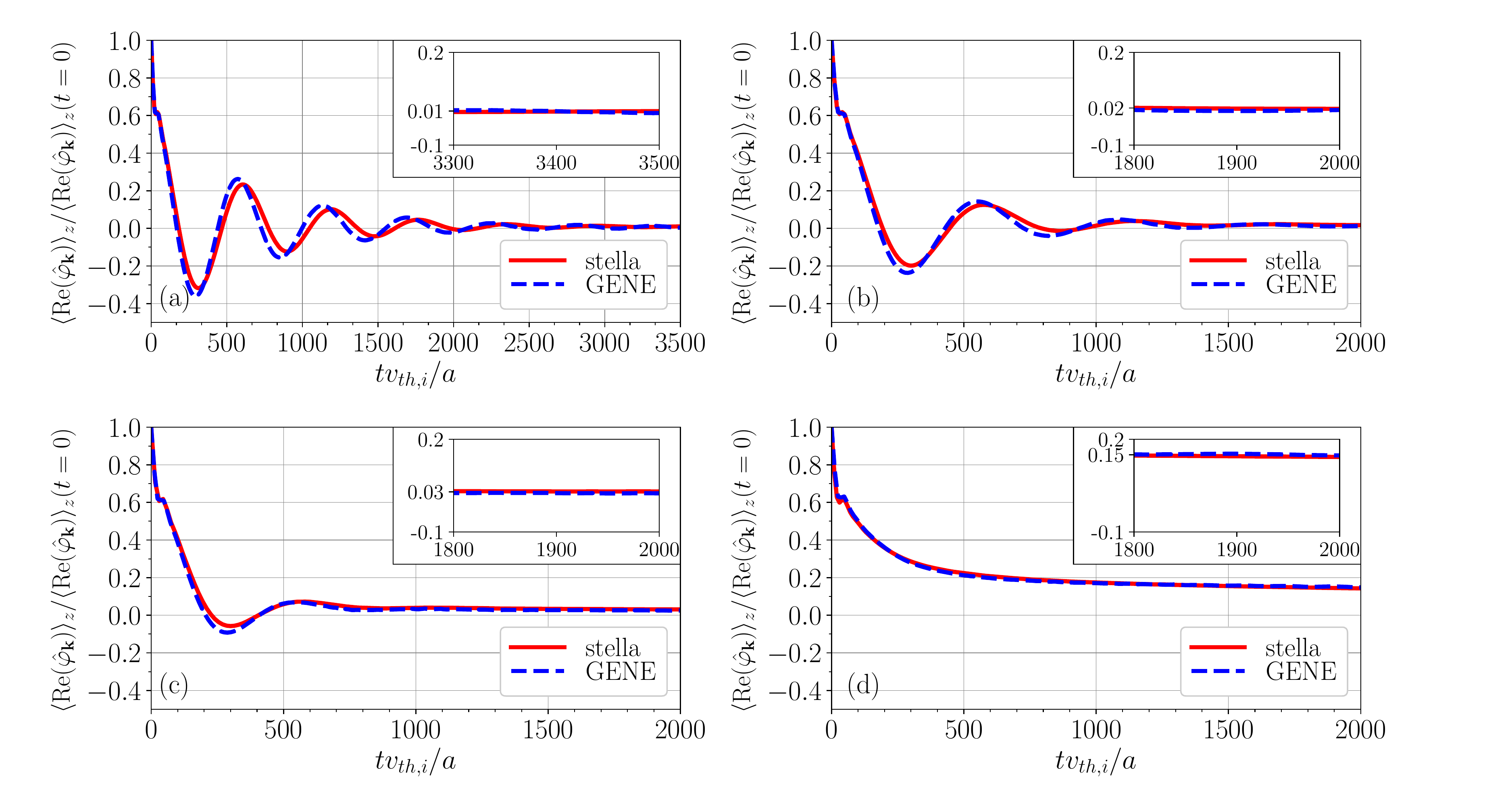}
    \caption{{For test 4, time} trace of the {line-averaged} electrostatic potential normalized to its maximum value computed with \texttt{stella} (solid red line) and with \texttt{GENE} (dashed blue line) for the pairs $(k_x\rho_i,k_y\rho_i)=(0.05,0)$ (a), $(k_x\rho_i,k_y\rho_i)=(0.07,0)$ (b), $(k_x\rho_i,k_y\rho_i)=(0.1,0)$ (c) and $(k_x\rho_i,k_y\rho_i)=(0.3,0)$ (d). The insets show a detail of each trace at large times.}
    \label{fig.ZF}
\end{figure}

For simplicity, in the simulations included in this test $a/L_{T_i}=a/L_{n_i}=0$ have been considered. {In addition,  as initial condition we have imposed $\hat{\varphi}_{\mathbf{k}}(t=0)$ to be a Gaussian function centered in the middle of the flux tube}.
Four time traces of the line average electrostatic potential normalized to its value at $t=0$ have been computed and represented in figures \ref{fig.ZF} (a)-(d). In these figures, it can be observed how the results obtained with both codes for $k_x\rho_i\in \{0.05, 0.07, 0.1, 0.3\}$ match remarkably well. As already mentioned, the four time traces show an initial GAM oscillation at $tv_{th,i}/a<100$, followed, except for figure \ref{fig.ZF} (d), by a lower-frequency damped oscillation. As observed in these figures, the frequency of the damped oscillation decreases with increasing $k_x$, in fact, for $k_x\rho_i=0.3$, represented in figure \ref{fig.ZF} (d), this frequency is missing. {The residual level of each time trace is given in the insets of these figures. These plots show that the residual value of the time traces increases with $k_x$.} 

%% file: files/Non_linear.tex
\section{Nonlinear simulations}\label{sec.5}

In flux tube nonlinear simulations, the codes solve equations (\ref{norm_gyro_eq}) and (\ref{quasi}) within a flux tube that extends in the radial and binormal directions. The parameters that define the flux tube for the nonlinear test presented in this section (test 5) are listed in table \ref{table_non_lin_param}. 

First nonlinear gyrokinetic simulations in W7-X were reported by \citet{Xanthopoulos2007PRL}, where \texttt{GENE} was used to study the nonlinear ITG-driven heat flux. Since then, \texttt{GENE} has been widely used to look at the nonlinear properties of turbulence in W7-X (\citet{Xanthopoulos2011}; \citet{Helander2015}). More recently, in \citet{BanNavarro2020}, the effects of ITG turbulent transport in different configurations of W7-X have been investigated with the global version of \texttt{GENE}, \texttt{GENE-3D} \citep{Maurer2020}. In \citep{Edi2020} the ITG-driven heat flux has been studied using realistic plasma parameters with the global particle-in-cell gyrokinectic code \texttt{EUTERPE}. Finally, simulations carried out with \texttt{stella} with all species treated kinetically have been employed to look at the transport of impurities driven by ITG and TEM turbulence in W7-X \citep{GarcaRegaa2021}.\\

\begin{table}
\centering
\begin{tabular}{c c c c c c c c c c c c c c c c c c c c c c c}
    \toprule
  Test 5. & &  $l_x/\rho_i$ & & & & $l_y/\rho_i$ & & & & $N_{k_x}$\footnote[1]{} & & & & $N_{k_y}$ & & & & $|k_{x}|_{\mathrm{min}}\rho_i$ & & & & $k_{y,\mathrm{min}}\rho_i$\\ \midrule
  \texttt{stella} & & 99.9 & & & & 62.8 & & & & 51 & & & & 64 & & & & 0.067 & & & & 0.100 \\
  \texttt{GENE} & & 131.3 & & & & 88.6 & & & & 101 & & & & 64 & & & & 0.047 & & & & 0.071 \\ \bottomrule

\end{tabular}
  \caption{Parameters used by \texttt{stella} and \texttt{GENE} to define the flux tube in test 5. From left to right: normalized flux tube size in the radial ($l_x$) and binormal ($l_y$) directions;  number of modes in the radial ($N_{k_x}$) and binormal ($N_{k_y}$) directions; smallest positive wavenumber in the radial (|$k_{x}|_{\mathrm{min}}$) and binormal ($k_{y,\mathrm{min}}$) directions.  }
  \label{table_non_lin_param}
\end{table}

\subsection{Test 5. Nonlinear ITG-driven heat flux}

\begin{figure}
    \centering
        \includegraphics[width=0.8\linewidth]{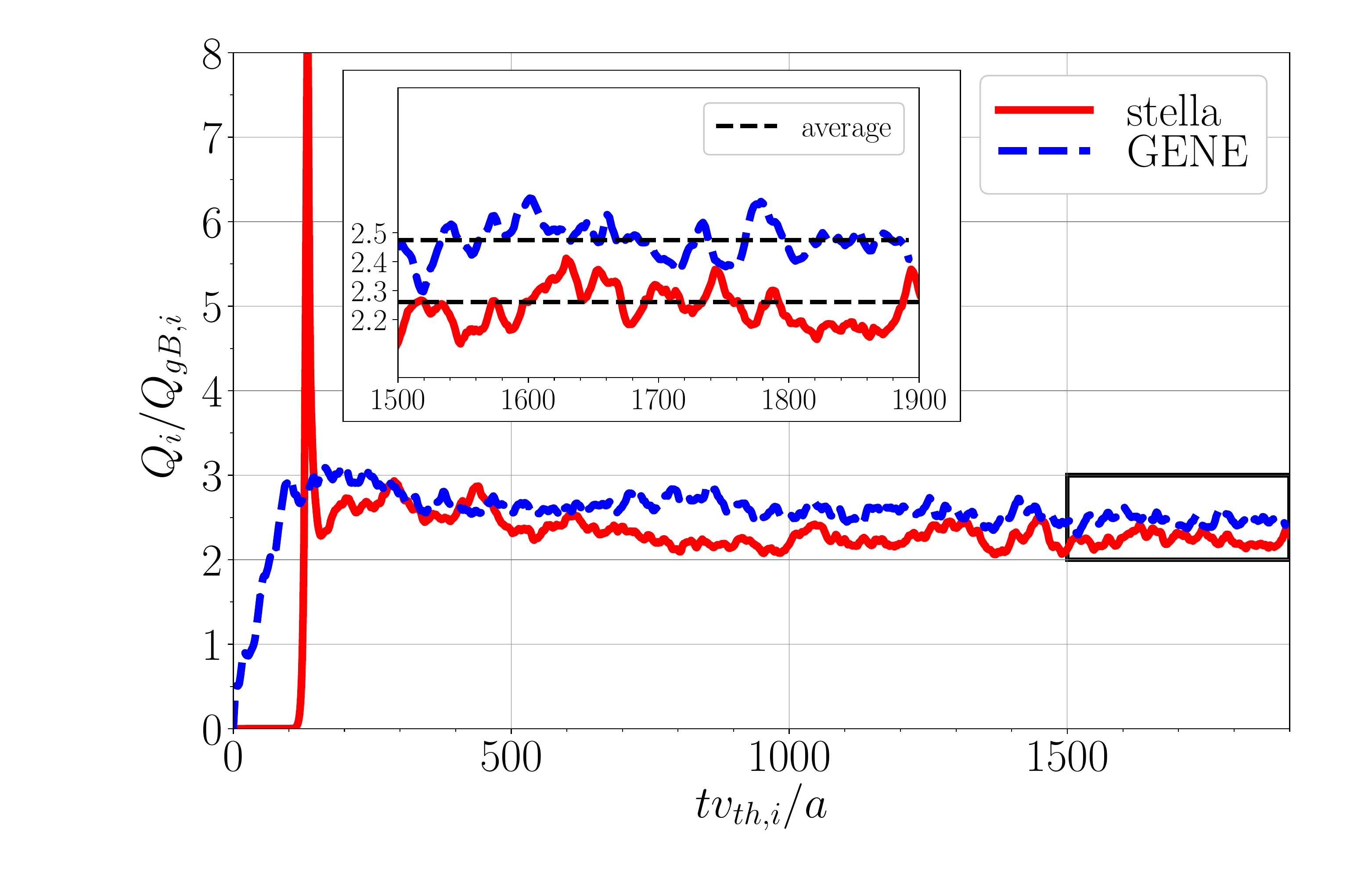}
    \caption{Time trace of the normalized ITG-driven heat flux computed with \texttt{stella} (solid red line) and \texttt{GENE} (dashed blue line) in the bean flux tube for the test 5 ITG scenario. The inset shows a detail of the time interval $tv_{th,i}/a=[1500, 1900]$, with the heat flux time average given by the black dashed lines.}
    \label{fig.non.lin}
\end{figure}

In this test, a nonlinear calculation of the ITG-driven heat flux, assuming adiabatic electrons, a normalized ion temperature gradient $a/L_{T_i}=3$ and a normalized ion density gradient $a/L_{n_i}=1$ is simulated (see tables \ref{table_param} and \ref{table_non_lin_param}\footnotetext[1]{$N_{k_x}$ includes both negative and positive values. $N_{k_y}$ only includes positive values.} for the values of the simulation parameters). The time trace of the ITG-driven heat flux computed with both codes and normalized to the ion gyro-Bohm heat flux, $Q_{gB,i}=n_iT_iv_{th_i}(\rho_i/a)^2$, is shown in figure \ref{fig.non.lin}. Despite the different initial evolution, both traces converge to very similar values. To quantify the difference between the saturated ITG-driven heat flux obtained with each code, an average over the time interval $tv_{th,i}/a=[1500, 1900]$ has been taken and represented in the inset of figure \ref{fig.non.lin}. The results for the normalized time averaged ITG-driven heat flux computed with \texttt{stella}, which is $Q_i/Q_{gB,i}=2.26$ and \texttt{GENE}, which is $Q_i/Q_{gB,i}=2.47$, represent a difference around $8.5\%$. This slight difference may be caused by the different resolution in the flux tube used by each code, see table \ref{table_non_lin_param}.

\begin{figure}
    \centering
    \includegraphics[width=0.55\linewidth]{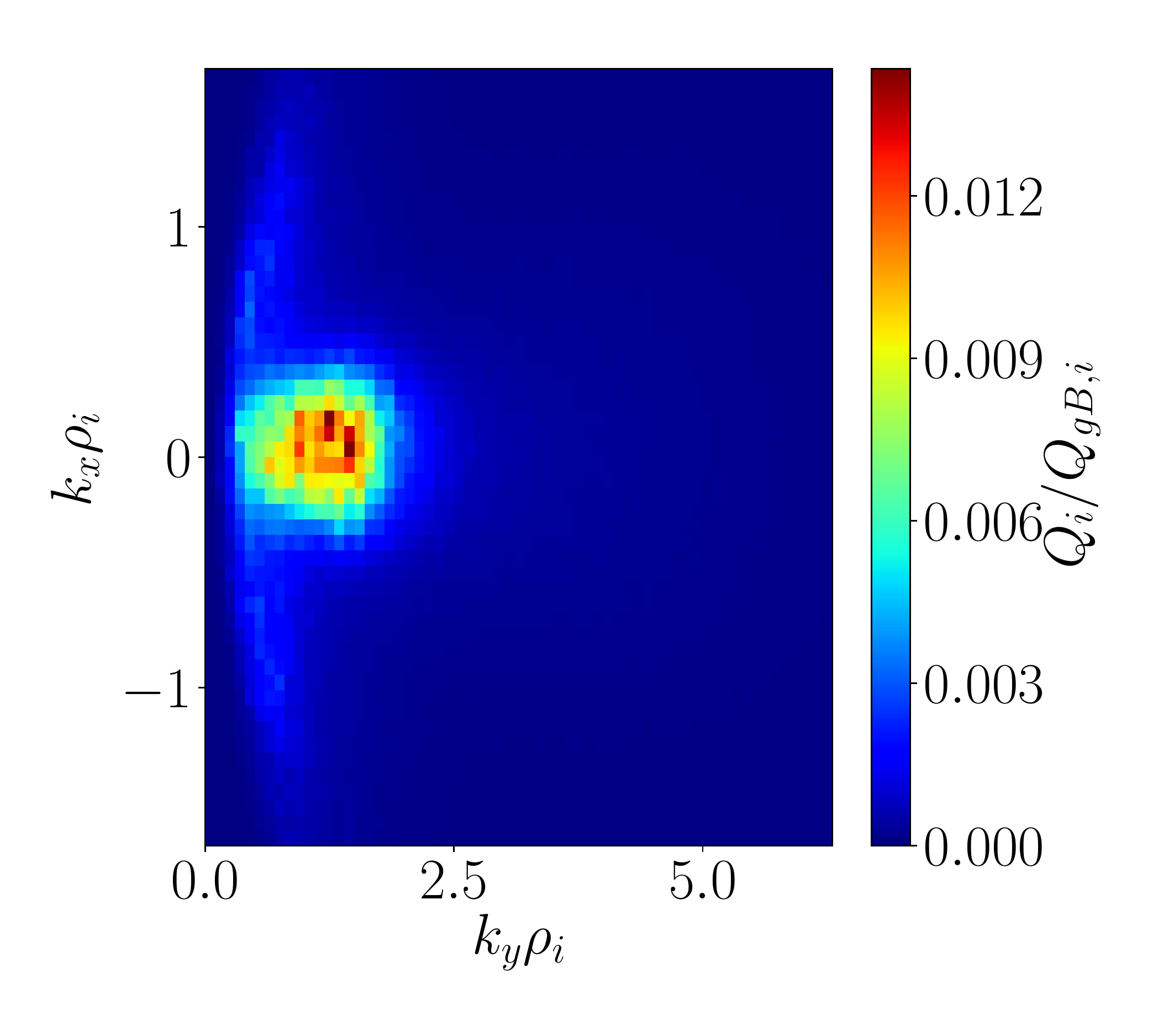}      
   \caption{Normalized nonlinear ITG-driven heat flux computed with \texttt{stella} in the bean flux tube, averaged over the time interval $tv_{th,i}/a=[1500, 1900]$ and represented as a function of $k_x$ and $k_y$.}
    \label{fig.Q_vs_kx_ky}
\end{figure}

\begin{figure}
    \centering
    \begin{subfigure}[b]{0.48\linewidth}        
        \centering
        \includegraphics[width=\linewidth]{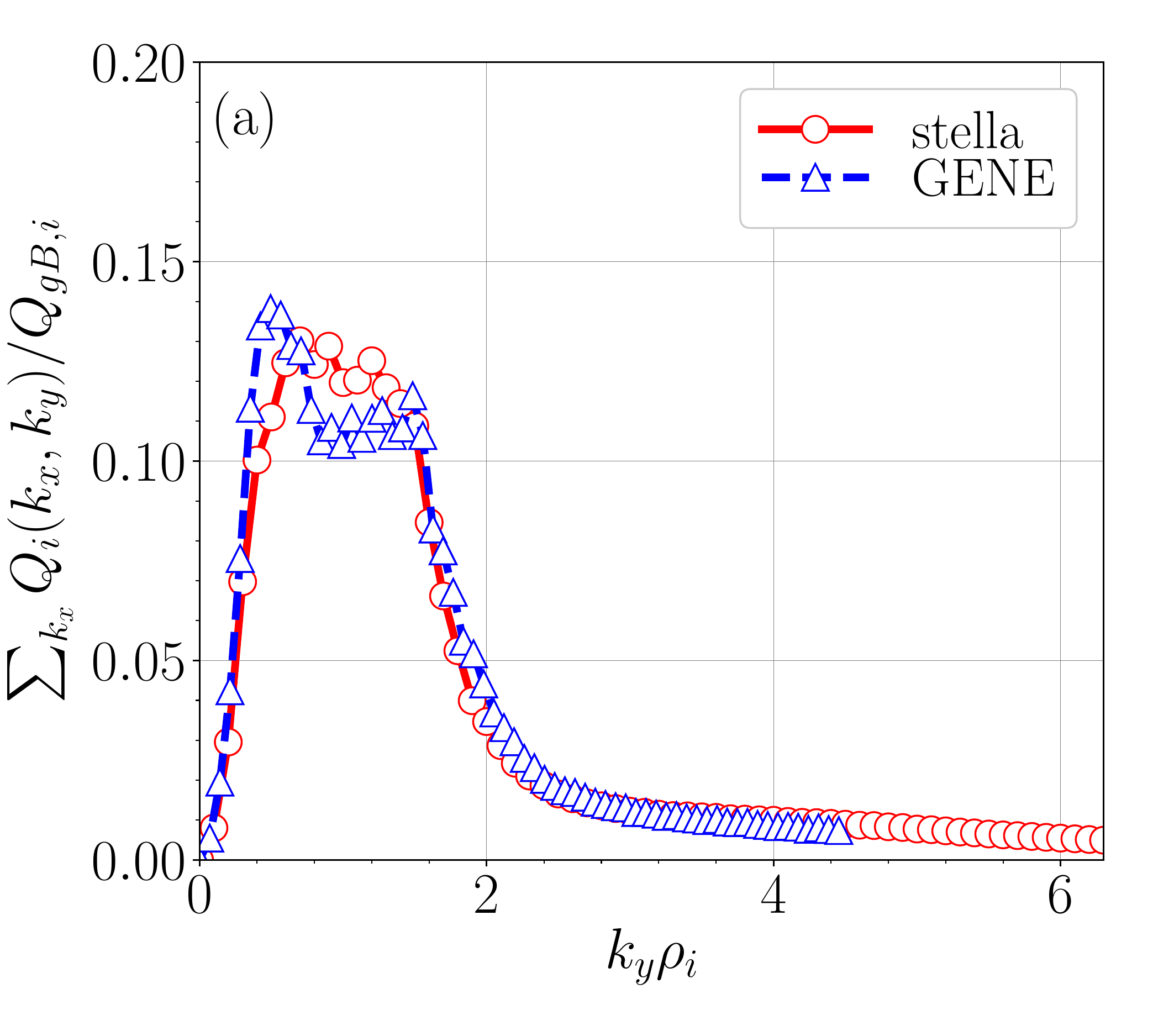}
    \end{subfigure}
\begin{subfigure}[b]{0.48\linewidth}        
        \centering
        \includegraphics[width=\linewidth]{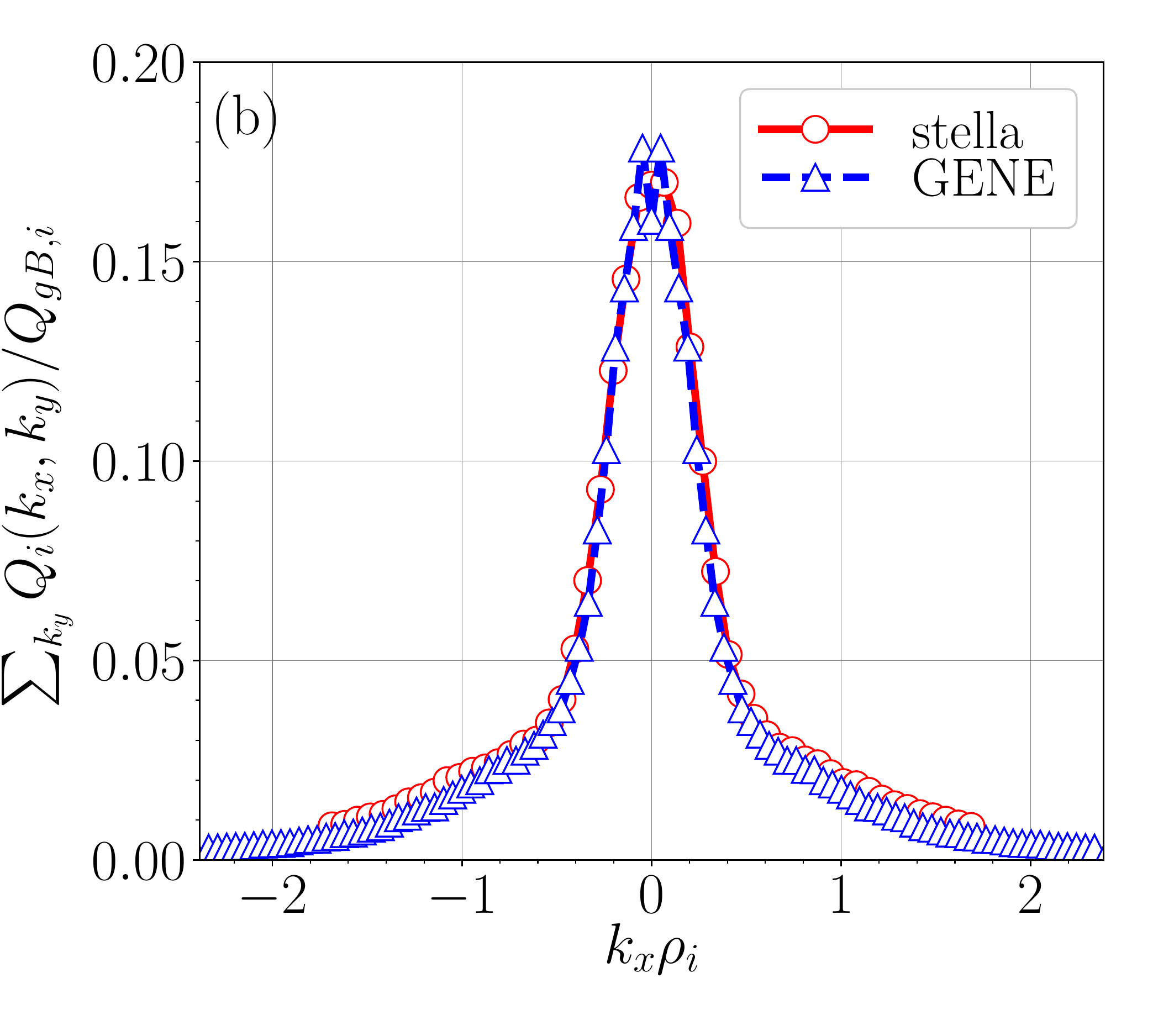}
    \end{subfigure}
    \caption{Normalized ITG-driven heat flux averaged over the time interval $tv_{th,i}/a=[1500, 1900]$ computed with \texttt{stella} (open circles linked by a solid red line) and \texttt{GENE} (open triangles linked by a dashed blue line) in the bean flux tube. It is represented as a function of $k_y$, summing over $k_x$ (a) and as a function of $k_x$, summing over $k_y$ (b).}
    \label{fig.Q_vs_k}
\end{figure}
In order to provide a more comprehensive study of these results, \texttt{stella} has been used to compute the contribution of each pair $(k_x,k_y)$ to the total ITG-driven heat flux. In figure \ref{fig.Q_vs_kx_ky}, it can be observed that the modes which contribute the most to the total heat flux are those with $k_y\rho_i\lesssim 2.0$ and $|k_x\rho_i|\lesssim 0.5$. To compare these results with \texttt{GENE} calculations, the ITG-driven heat flux is represented as a function of $k_y$, summing over $k_x$, in figure \ref{fig.Q_vs_k} (a), and as function of $k_x$, summing over $k_y$, in figure \ref{fig.Q_vs_k} (b). These figures show a satisfactory agreement between both codes.

\begin{figure}
    \centering
    \begin{subfigure}[b]{0.48\linewidth}        
        \centering
        \includegraphics[width=\linewidth]{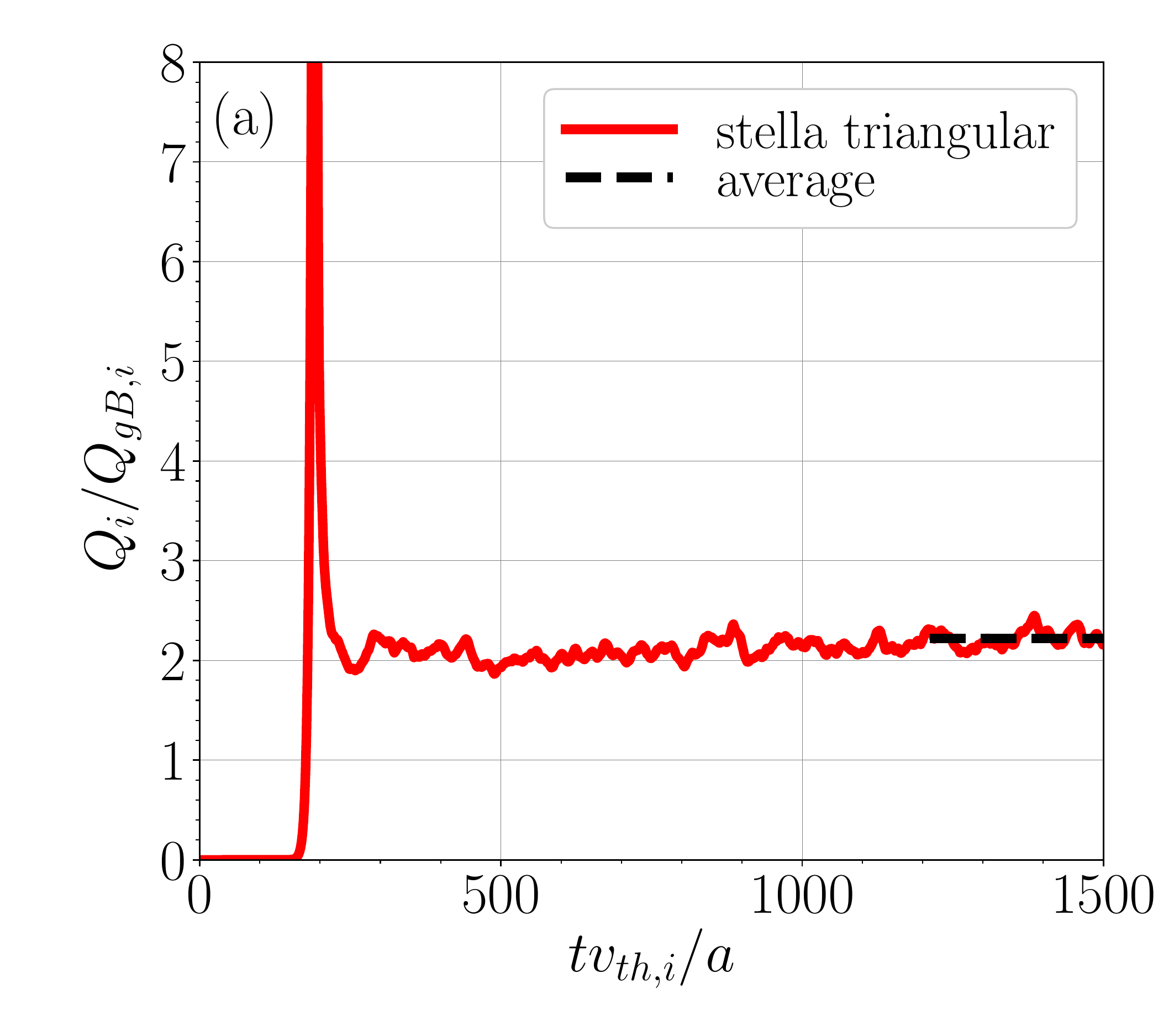}
    \end{subfigure}
\begin{subfigure}[b]{0.48\linewidth}        
        \centering
        \includegraphics[width=\linewidth]{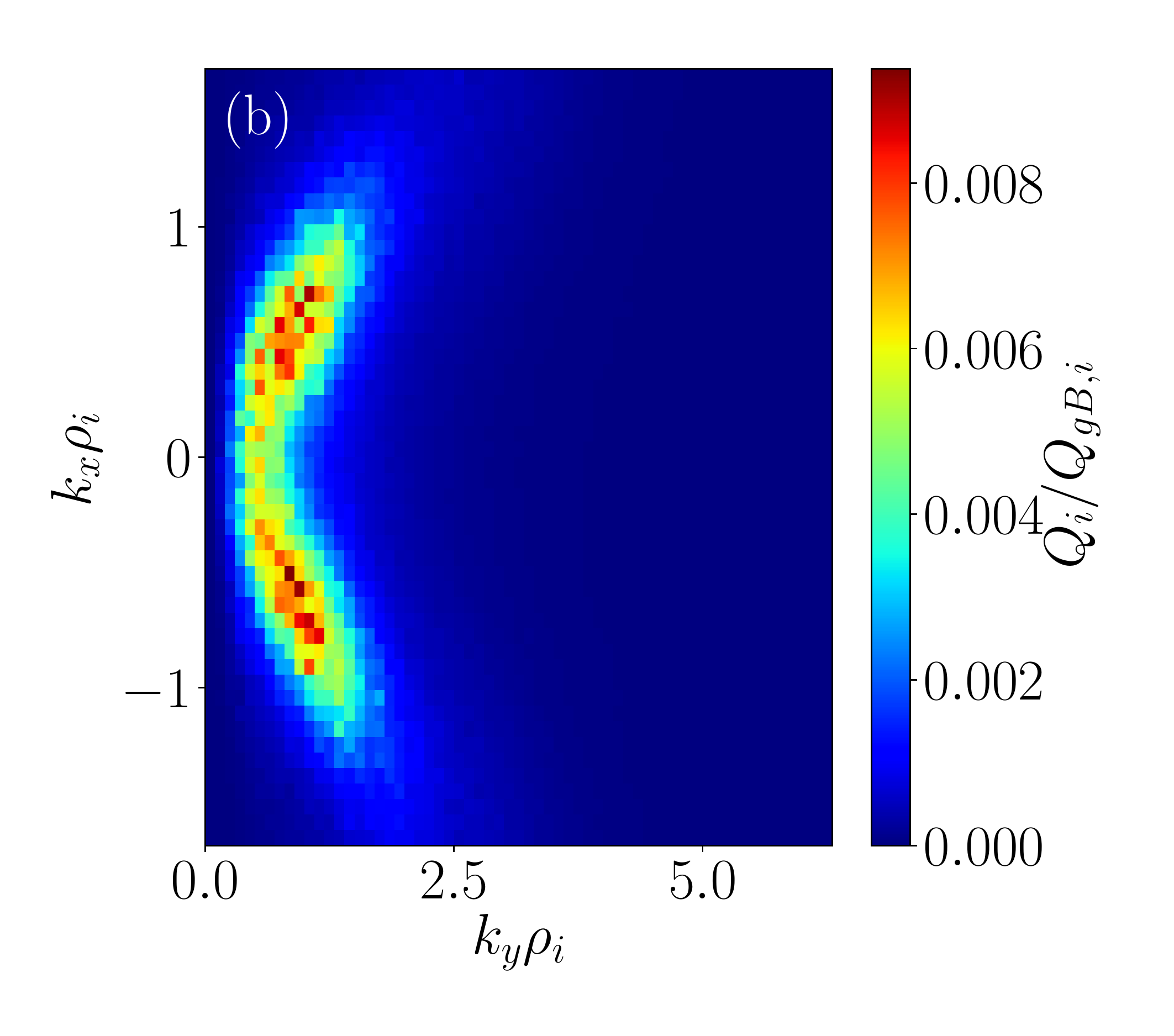}
    \end{subfigure}
    \caption{Time trace of the normalized ITG-driven heat flux computed with \texttt{stella} in the triangular flux tube, together with the heat flux time average over $tv_{th,i}/a=[1200,1500]$, represented with a black dashed line (a). This averaged ITG-driven heat flux is also represented as a function of $k_x$ and $k_y$ (b).}
    \label{fig.non_lin_triangular}
\end{figure}

Finally, for the sake of completeness, a simulation in the triangular flux tube performed with \texttt{stella} has been included in this section. The parameters selected to carry this simulation out are the ones collected in tables \ref{table_param} and \ref{table_non_lin_param} for test 5. The time trace of the normalized ITG-driven heat flux and the contribution of each mode are represented in figures \ref{fig.non_lin_triangular} (a) and \ref{fig.non_lin_triangular} (b), respectively. Figure \ref{fig.non_lin_triangular} (a) shows the saturated ITG-driven heat flux in the triangular flux tube to be $Q_i/Q_{gB,i}=2.22$. This value is very similar to the one obtained in the bean flux tube using \texttt{stella}. The main differences between nonlinear ITG-driven heat flux calculations in both flux tubes are found in their spectrum, as it can be seen by comparing the maps given in figures \ref{fig.Q_vs_kx_ky} and \ref{fig.non_lin_triangular} (b). As in the bean flux tube, the modes in the binormal direction which contribute the most to the total heat flux in the triangular flux tube are those with $k_y\rho_i\lesssim 2.0$. However, in contrast with the results obtained for the bean flux tube, modes in the radial direction with $0.5\lesssim|k_x\rho_i|\lesssim 1.5$ give a large contribution to the total heat flux in the triangular flux tube.

%% file: files/Summary.tex
\section{Summary and conclusions}\label{sec.6}
{Due to the increasing interest in stellarator gyrokinetic modelling, fostered by the results of W7-X first experimental campaigns, it is desirable to have a sufficiently complete, documented and well verified set of linear and nonlinear gyrokinetic simulations in W7-X geometry against which present and future stellarator gyrokinetic codes can be tested and benchmarked. In this paper, such a set of simulations has been provided in the form of a comprehensive benchmark between the codes \texttt{stella} and \texttt{GENE}. This benchmark, consisting of five different tests, has been carried out in a fixed-boundary high-mirror configuration of W7-X. The linear part of the benchmark has been} presented in tests 1 to 4. ITG instabilities have been studied in the bean and triangular flux tubes of W7-X in tests 1 and 2, respectively. Comparing these results, it can be concluded that both flux tubes are equally unstable, but the largest growth rates are found at different radial wavenumbers. TEM instabilities {driven by density gradients} have been studied in test 3, where it has been shown how the mixed implicit-explicit method used by \texttt{stella} allows to use larger time steps than explicit codes for simulations with kinetic electrons. In these three tests, the structure of the electrostatic potential associated {with} each instability has been {given} and the growth rate and real frequency values obtained with \texttt{stella} and \texttt{GENE} have been successfully compared. In test 4, different time traces of the zonal electrostatic potential relaxation have been compared. Finally, the nonlinear ITG-driven heat flux and its spectrum have been calculated in test 5 in the bean flux tube with both codes and, for completeness, in the triangular flux tube with \texttt{stella}. The computed energy fluxes are similar in both flux tubes, although the radial modes that give the largest contribution to the total heat flux are different.

%% file: files/Acknow.tex
\section*{Acknowledgments}

This work has been carried out within the framework of the EUROfusion Consortium
and has received funding from the Euratom research and training programme 2014-2018
and 2019-2020 under grant agreement No. 633053. The views and opinions expressed
herein do not necessarily reflect those of the European Commission. This research was
supported in part by grant PGC2018-095307-B-I00, Ministerio de Ciencia, Innovación y
Universidades, Spain. The simulations were carried out in the clusters Marconi (Cineca,
Italy) and Xula (Ciemat, Spain). J.M.G.R and A.G.J acknowledge the hospitality of the Max-Planck Institut für Plasmaphysik, Greifswald, and of the Rudolf Peierls Centre for Theoretical Physics, University of Oxford, where part of this work was done. A.G.J thanks R. Jorge and M. Landreman (University of Maryland) for their advice on the visualization of the VMEC equilibria.

%% file: files/AP_1.tex
\section{VMEC parameters to reproduce the studied equilibria}\label{aped.A}
In this appendix, the list of input parameters used to generate the fixed-boundary VMEC equilibrium used in the present work is provided (see the beginning of section \ref{sec.parameters}):

\begin{lstlisting}
&INDATA
 MGRID_FILE = ''
 LFREEB     = F 
 LWOUTTXT   = T
 LDIAGNO    = T
 DELT       = 0.9
 TCON0      = 2.
 NFP        = 5
 NCURR      = 1
 MPOL       = 12      NTOR =  12
 NZETA      = 36
 NS_ARRAY   = 9  28  99
 NITER      = 3700
 NSTEP      = 100
 NVACSKIP   = 6
 GAMMA      = 0.
 FTOL_ARRAY = 0. 1.E-10 1.E-14
 PHIEDGE    = 2.0
 BLOAT      = 1.
 CURTOR     = 0.
 SPRES_PED  = 1.
 AM         = 179370 -179370
 AI         = 0.36 -0.062 0.0712 -0.025 0. 0. 0. 0. 0. 0. 0.
 AC         = 0. 0. 0. 0. 0. 0. 0. 0. 0. 0.
 RAXIS      = 5.5395 0.37547 0.01698 0.0014216 0.000027606
 ZAXIS      = 0. -0.31289 -0.019191 -0.00044262 0.000034742
  RBC( 0, 0)= 5.5021E+00  ZBS( 0, 0)= 0.0000E+00
  RBC( 1, 0)= 2.8455E-01  ZBS( 1, 0)=-2.4009E-01
  RBC( 2, 0)=-5.7247E-03  ZBS( 2, 0)= 1.9753E-03
  RBC( 3, 0)=-3.4624E-04  ZBS( 3, 0)= 2.2597E-03
  RBC( 4, 0)=-1.4722E-03  ZBS( 4, 0)= 1.7290E-03
  RBC(-4, 1)=-7.8450E-04  ZBS(-4, 1)=-7.6096E-06
  RBC(-3, 1)=-1.4587E-04  ZBS(-3, 1)= 1.2663E-03
  RBC(-2, 1)= 1.5679E-03  ZBS(-2, 1)= 6.9395E-03
  RBC(-1, 1)= 1.8726E-02  ZBS(-1, 1)= 1.6401E-02
  RBC( 0, 1)= 4.8197E-01  ZBS( 0, 1)= 6.0460E-01
  RBC( 1, 1)=-2.1345E-01  ZBS( 1, 1)= 1.9431E-01
  RBC( 2, 1)=-2.1789E-02  ZBS( 2, 1)= 1.8594E-02
  RBC( 3, 1)= 1.4488E-03  ZBS( 3, 1)=-1.1209E-03
  RBC( 4, 1)= 1.4451E-03  ZBS( 4, 1)=-7.5419E-04
  RBC(-4, 2)= 1.4744E-04  ZBS(-4, 2)= 1.5260E-04
  RBC(-3, 2)= 2.9997E-04  ZBS(-3, 2)=-2.6406E-05
  RBC(-2, 2)= 2.0710E-03  ZBS(-2, 2)=-2.4917E-05
  RBC(-1, 2)= 1.0333E-02  ZBS(-1, 2)= 7.7931E-03
  RBC( 0, 2)= 3.5672E-02  ZBS( 0, 2)=-2.0312E-03
  RBC( 1, 2)= 4.3441E-02  ZBS( 1, 2)= 1.9773E-02
  RBC( 2, 2)= 6.6787E-02  ZBS( 2, 2)=-4.9779E-02
  RBC( 3, 2)=-2.1551E-04  ZBS( 3, 2)= 1.6853E-03
  RBC( 4, 2)=-5.9507E-04  ZBS( 4, 2)= 7.7889E-04
  RBC(-4, 3)= 1.5743E-04  ZBS(-4, 3)=-2.1339E-04
  RBC(-3, 3)=-1.4708E-04  ZBS(-3, 3)= 3.1735E-06
  RBC(-2, 3)=-1.2572E-04  ZBS(-2, 3)= 6.1239E-04
  RBC(-1, 3)= 1.5997E-03  ZBS(-1, 3)=-2.9522E-04
  RBC( 0, 3)=-2.0732E-03  ZBS( 0, 3)=-1.8327E-03
  RBC( 1, 3)=-1.1519E-02  ZBS( 1, 3)=-4.6829E-03
  RBC( 2, 3)=-2.0190E-02  ZBS( 2, 3)= 6.7910E-03
  RBC( 3, 3)=-1.3317E-02  ZBS( 3, 3)= 1.1217E-02
  RBC( 4, 3)= 1.2037E-03  ZBS( 4, 3)=-8.0214E-04
  RBC(-4, 4)=-3.6955E-06  ZBS(-4, 4)= 1.2264E-04
  RBC(-3, 4)= 1.9747E-04  ZBS(-3, 4)=-1.7509E-04
  RBC(-2, 4)= 9.2916E-05  ZBS(-2, 4)= 6.5936E-05
  RBC(-1, 4)=-4.2286E-04  ZBS(-1, 4)= 2.4530E-05
  RBC( 0, 4)= 1.9963E-03  ZBS( 0, 4)= 4.5823E-04
  RBC( 1, 4)=-1.2553E-03  ZBS( 1, 4)= 1.7370E-03
  RBC( 2, 4)= 8.1636E-03  ZBS( 2, 4)= 8.8024E-03
  RBC( 3, 4)= 3.0460E-03  ZBS( 3, 4)=-4.6653E-03
  RBC( 4, 4)= 5.0489E-04  ZBS( 4, 4)=-1.5253E-03
\end{lstlisting}